\documentclass[preprint,11pt]{elsarticle}
\journal{Computer Physics Communications}

\usepackage[title,titletoc]{appendix}
\usepackage{amssymb}
\usepackage{amsmath}
\usepackage{graphicx}
\usepackage{subfigure}
\usepackage{multirow}
\usepackage{makecell}
\usepackage{bm}
\usepackage{color}
\usepackage{appendix}
\usepackage{fancyhdr}
\usepackage[utf8]{inputenc}
\usepackage{caption}
\usepackage[font=footnotesize]{caption}
\DeclareCaptionFont{white}{\color{white}}
\DeclareCaptionFormat{listing}{\colorbox{gray}{\parbox{\textwidth}{#1#2#3}}}

\biboptions{sort&compress}

\newcounter{bla}

\newcommand{\be}{\begin{equation}}
\newcommand{\ee}{\end{equation}}
\newcommand{\bs}{\begin{subequations}}
\newcommand{\es}{\end{subequations}}
\newcommand{\beal}{\begin{align}}
\newcommand{\munu}{{\mu\nu}}
\newcommand{\eq}{\mathrm{eq}}

\newcommand{\LRF}{\mathrm{LRF}}
\newcommand{\ene}{\mathcal{E}}
\newcommand{\Peq}{\mathcal{P}_\eq}
\newcommand{\I}{{\sc iS3D}}

\newcommand{\U}{{\sc UrQMD}}
\newcommand{\bp}{{\bm p}}
\newcommand{\A}{\mathcal{A}}
\newcommand{\B}{\mathcal{B}}
\newcommand{\N}{\mathcal{N}}
\newcommand{\J}{\mathcal{J}}

\definecolor{olive}{rgb}{0.5, 0.5, 0.0}
\definecolor{byzantium}{rgb}{0.44, 0.16, 0.39}
\definecolor{dartmouthgreen}{rgb}{0.05, 0.5, 0.06}

\begin{document}
\begin{frontmatter}

\title{{\bf Particlization in fluid dynamical simulations of heavy-ion collisions: The \I\ module}}

\author[a]{Mike McNelis\corref{corauthor}}
\author[a]{Derek Everett\corref{corauthor}}
\author[a,c]{Ulrich Heinz}
\cortext[corauthor] {Corresponding author.\\ Email addresses: mcnelis.9@osu.edu (McNelis), everett.165@osu.edu (Everett), heinz.9@osu.edu (Heinz)}
\address[a]{Department of Physics, The Ohio State University, Columbus, OH 43210-1117, USA}
\address[c]{Institut f\"ur Theoretische Physik, J.~W.~Goethe Universit\"at, Max-von-Laue-Str. 1, D-60438 Frankfurt am Main, Germany}
\date{\today}

\begin{abstract}
    The \I\ particlization module simulates the emission of hadrons from heavy-ion collisions via Monte-Carlo sampling of the Cooper-Frye formula which converts fluid dynamical information into  local phase-space distributions for hadrons. The code package includes multiple choices for the non-equilibrium correction to these distribution functions: the 14-moment approximation, first-order Chapman-Enskog expansion, and two types of modified equilibrium distributions. This makes it possible to explore to what extent heavy-ion experimental data are sensitive to different choices for $\delta f_n$, presently the main source of theoretical uncertainty in the particlization stage. We validate our particle sampler with a high degree of precision by generating several million hadron emission events from a longitudinally boost-invariant hypersurface and comparing the event-averaged particle spectra and space-time distributions to the Cooper-Frye formula.
\end{abstract}    


\end{frontmatter}

\newpage

\noindent
{\bf PROGRAM SUMMARY}
\vspace{0.5cm}\\
\begin{small}
\textit{Manuscript Title:} Particlization in fluid dynamical simulations of heavy-ion collisions: The \I\ module\\ 
\textit{Authors:} Mike McNelis, Derek Everett, Ulrich Heinz \\
\textit{Program Title:} \I\ \\
\textit{Journal Reference:} \\
\textit{Catalogue Identifier:} \\
\textit{Licensing provisions:} GPLv3 \\
\textit{Programming Language:} C++, CUDA C \\
\textit{Computer:} Laptop, desktop, cluster (an Nvidia graphics processing unit is optional) \\
\textit{Operating System:} GNU/Linux distributions \\
\textit{Global memory usage:} 1 GB (to sample 1000 events from a 2+1d hypersurface) 
\textit{Classification:} 12 Gases and Fluids, 17 Nuclear physics \\
\textit{External routines/libraries:} GNU Scientific Library (GSL) \\
\textit{Nature of problem:\\}
Sampling the emission of hadrons during the particlization stage of heavy-ion collisions and accelerating the continuous Cooper-Frye formula. \\
\textit{Solution method:\\}
Monte-Carlo simulation, parallel computing \\
\textit{Running time:\\}
The typical running time to sample 1000 particlization events is about 94$s$ on a single-core Intel Xeon E5-2680 v4 CPU. The time it takes to compute the continuous momentum spectra on a $p_T \times \phi_p = 100 \times 48$ grid is about 56$s$ per particle on an Intel Xeon E5-2680 v4 multi-core processor with OpenMP and 2.1$s$ per particle on an Nvidia Tesla V100-PCIE graphics card. These benchmark tests are done taking the 14-moment approximation for the $\delta f_n$ correction and using a hypersurface from a longitudinally boost-invariant central Pb-Pb collision at LHC energies, which contains approximately $1.9 \times 10^5$ freezeout cells.
\end{small}

\newpage
\tableofcontents
\newpage


\section{Introduction}
\label{S1}

In the hadronization phase of a heavy-ion collision, the quark-gluon plasma undergoes a phase transition into a hadron resonance gas. The conversion of strongly coupled partonic degrees of freedom, modeled by fluid dynamics, into weakly coupled hadronic degrees of freedom is possible under the assumption that hydrodynamics and kinetic theory have an overlapping region of validity on a hypersurface $\Sigma$. On $\Sigma$, one can compute the momentum spectrum of each hadron species $n$ using the Cooper-Frye formula \cite{PhysRevD.10.186}
\be
\label{eq1}
  E_p \frac{dN_n}{d^3p} = \frac{1}{(2\pi\hbar)^3}\int_\Sigma p \cdot d^3\sigma(x) \, f_n(x,p) \,,
\ee
where $f_n(x,p)$ is the phase-space distribution function for the given hadron species. If the fluid is in local thermal and chemical equilibrium, $f_n$ reduces to the Maxwell-J\"{u}ttner distribution
\begin{equation}
\label{eq:feq}
  f_{\eq,n} (x,p) = g_n 
  \left[\exp
         \left(\frac{p \cdot u(x)}{T(x)} - b_n \alpha_B(x) \right) + \Theta_n
  \right]^{-1},
\end{equation}
where $g_n=2s_n{+}1$ and $b_n$ are the spin degeneracy and baryon number of species $n$, $u^{\mu}(x)$ is the fluid velocity, $T(x)$ is the temperature, $\alpha_B(x) = \mu_B(x)/T(x)$ is the ratio of baryon chemical potential and temperature, and $\Theta_n = {(-1,1)}$ accounts for the quantum statistics of the particles (Bose-Einstein or Fermi-Dirac).\footnote{%
    Different isospin (electric charge) and strangeness states within a hadronic multiplet (e.g.\ $K^+$ and $K^0$) are counted as separate hadron species; however, we here ignore chemical potentials associated with these additional conserved charges.}
%
%
However, viscous and diffusive fluids are generally out of local equilibrium, and $f_n$ will therefore have a non-equilibrium correction $\delta f_n$:
\be
\label{eq-decomp}
    f_n(x,p) = f_{\eq,n}(x,p) + \delta f_n(x,p).
\ee
Some knowledge about $\delta f_n$ comes from the net baryon current $J_B^{\mu}$ and the energy-momentum tensor $T^{\mu\nu}$ provided by the preceding viscous hydrodynamic evolution. In kinetic theory, $J_B^\mu$ and $T^{\mu\nu}$ are the first and second moments of the distribution function, respectively:
\begin{equation}
\label{eq:JmuTmunu} 
 J_B^{\mu}(x) = \sum_n b_n  \int_p p^{\mu} f_n(x,p),\qquad
 T^{\mu\nu}(x) = \sum_{n} \int_p p^{\mu}p^{\nu} f_n(x,p),
\end{equation}
where $\int_p \equiv \int{\dfrac{d^3p}{(2\pi\hbar)^3E_p}}$ and the sum over $n$ goes over all $N_R$ different hadron resonance species.\footnote{%
    Different isospin (charge) states within a hadronic multiplet (e.g. $\pi^+$, $\pi^0$, $\pi^-$) are counted as separate hadron species.}
The largest contributions to these hydrodynamic moments come from particles with {\it thermal} momenta $p\sim T$, i.e. they are not sensitive to details of the highly suppressed large-momentum tails of the distribution function. Still, using the decomposition (\ref{eq-decomp}), the constraints (\ref{eq:JmuTmunu}) allow for an infinity of choices for the momentum dependence of $\delta f_n(x,p)$; this renders the transition from fluid to particles intrinsically ambiguous. In particular, the hydrodynamic output provides basically no useful information about hadrons emitted with high (i.e. much larger than thermal) momenta; to predict the final distributions of such \textit{hard} particles requires coupling the macroscopic hydrodynamic evolution of the soft medium to a full microscopic kinetic theory for them (see, for example,  \cite{Akamatsu:2018olk, Pierog:2013ria}). 

We here focus only on those parts of the local momentum distribution that contribute significantly to the hydrodynamic moments. To constrain the form of $f_n$ beyond what is required by Eqs.~(\ref{eq:JmuTmunu}) we will assume the simultaneous applicability of dissipative fluid dynamics and classical relativistic kinetic theory on the conversion surface and therefore develop ans\"atze for $\delta f_n$ that are compatible with $f_n$ solving the Boltzmann equation
\begin{equation}
\label{eq:boltzmann}
  p^\mu \partial_\mu f_n(x,p) = \mathcal{C}_n(x,p)\,,
\end{equation}
with some reasonable but simple approximate form of the collision term $\mathcal{C}_n(x,p)$.\footnote{%
    Accounting in the particlization algorithm for the
    full complexity of allowed collision processes among
    the different hadron species and for dynamical consistency of $\delta f_n$ with a realistic collision term in the Boltzmann equation (\ref{eq:boltzmann}) is numerically quite challenging \cite{Molnar:2014fva, Wolff:2016vcm, Damodaran:2017ior}.}
A common assumption is to linearize $\delta f_n$ in the shear stress tensor $\pi^{\mu\nu}$, bulk viscous pressure $\Pi$ and baryon diffusion current $V_B^{\mu}$:
\be
\begin{split}
\label{eq:df_linear}
  \delta f_n(x,p) \approx& \, c_{\pi,n}(x, p) \, 
  p_{\langle\mu} p_{\nu\rangle}\,\pi^{\mu\nu}(x) 
  + c_{\Pi,n}(x, p)\,\Pi(x) \\ 
  &+ c_{V,n}(x, p)\,p_{\langle\mu\rangle}V_B^{\mu}(x)\,,
  \end{split}
\ee
where $c_{\pi,n}$, $c_{\Pi,n}$ and $c_{V,n}$ are expansion coefficients.\footnote{%
    Note that the expansion coefficients have both spatial and momentum dependence through $f_{\eq,n}(x,p)$, $T(x)$, $\alpha_B(x)$, and powers of $u(x)\cdot p$.}$^,$\footnote{%
    Angular brackets around one Lorentz index indicate projection of a vector onto its locally spatial part, $A^{\langle\mu\rangle}\equiv\Delta^{\mu}_{\alpha} A^{\alpha}$, with $\Delta^\mu_\alpha{\,=\,}g^\mu_\alpha{-} u^\mu u_\alpha$ projecting onto the spatial directions in the fluid's local rest frame (LRF) defined by the flow velocity $u^\mu$. Angular brackets around a pair of Lorentz indices denote the projection of a tensor onto its traceless and locally spatial part, $A^{\langle\mu\nu\rangle} \equiv \Delta^{\mu\nu}_{\alpha\beta} A^{\alpha\beta}$, with $\Delta^{\mu\nu}_{\alpha\beta} = \frac{1}{2}(\Delta^\mu_\alpha\Delta^\nu_\beta{+} \Delta^\mu_\beta \Delta^\nu_\alpha) - \frac{1}{3} \Delta^{\mu\nu} \Delta_{\alpha\beta}$.}
Two common choices for $\delta f_n$, the 14-moment approximation \cite{CPA:CPA3160020403} and the first-order Chapman-Enskog expansion \cite{chapman1990mathematical}, are based on this assumption. Recently developed techniques for computing $\delta f_n$ do not linearize the expansion around $f_{\eq,n}$ but rather apply viscous corrections directly to the exponent of the Boltzmann factor in Eq.\,\eqref{eq:feq}. The particlization model \I\ described in this work also provides options to use two variants of such \textit{modified equilibrium distributions}: the Pratt-Torrieri-Bernhard (PTB) distribution \cite{Pratt:2010jt, Bernhard:2018hnz} and a new variant of this idea which we call the PTM distribution \cite{feqmod}; both will be briefly reviewed below.

Recently, much effort has gone into developing tools to extract medium properties of the quark-gluon plasma (QGP), including the transport coefficients describing viscous and diffusive effects within the QGP fluid itself and energy and momentum exchange processes between this fluid and hard probes (light and heavy flavor jets) travelling through it, from a global comparison of dynamical models with a large set of available experimental data using modern Bayesian statistical methods \cite{Bernhard:2016tnd, Putschke:2019yrg}. It is important to understand that the final hadronic observables depend on the specific shear and bulk viscosities $\eta/\mathcal{S}$ and $\zeta/\mathcal{S}$ (where $\mathcal{S}$ is the entropy density) and the baryon diffusion coefficient $\kappa_B$ in two ways: {\sl (i)} through the transport equations for the shear stress $\pi^{\mu\nu}$, bulk viscous pressure $\Pi$ and baryon diffusion current $V_B^{\mu}$, which modify the ideal hydrodynamic evolution of the temperature $T(x)$, flow velocity $u^\mu(x)$ and baryon chemical potential $\alpha_B(x)$ of the fluid which all enter in the exponent of the equilibrium part of the distribution (\ref{eq:feq}), and {\sl (ii)} directly through the values of $\pi^{\mu\nu}(x)$, $\Pi(x)$ and $V_B^{\mu}(x)$ on the particlization surface which enter the dissipative corrections $\delta f_n$. The first effect {\sl (i)} records the history of the transport coefficients $\eta/\mathcal{S}$, $\zeta/\mathcal{S}$, and $\kappa_B$ integrated over the entire evolution of the fluid whereas the second effect {\sl (ii)} is (up to short-term memory effects limited by the microscopic relaxation time) only sensitive to the values of these coefficients directly on the particlization surface. For a given set of initial conditions, hydrodynamic evolution equations and transport coefficients\footnote{%
    For example, the  hydrodynamic code MUSIC \cite{Schenke:2010nt} uses the Denicol-Niemi-M\'olnar-Rischke (DNMR) $2^{\text{nd}}$-order relaxation equations \cite{Denicol:2012cn} to evolve $\pi^{\mu\nu}$, $\Pi$ and $V_B^\mu$.}
the effect {\sl (i)} is fixed but the effect {\sl (ii)} on the emitted particle spectra is still plagued by ambiguities related to the choice of parametrization of $\delta f_n$ in terms of $\pi^{\mu\nu}$, $\Pi$ and $V_B^\mu$. \I\ is a numerical tool that allows one to explore the phenomenological consequences of these ambiguities and how they propagate into theoretical uncertainties of the medium parameters extracted from a Bayesian analysis of experimental data.

\I\ is a C++ particlization code that was developed from the code {\sc iSS} written for the {\sc iEBE-VISHNU} dynamical simulation code package \cite{Shen:2014vra}, by extending it to 3-dimensionally expanding systems without longitudinal boost-invariance and adding additional options for the form of the distribution functions $f_n(x,p)$.\footnote{%
    The \I\ code package is publicly available for download from the GitHub repository \url{https://github.com/derekeverett/iS3D}.
    \label{fn7}}
It allows for computing and sampling the Cooper-Frye formula using one of four choices for $\delta f_n$: the 14-moment approximation, the first-order Chapman-Enskog expansion, and the PTB and PTM modified equilibrium distributions. The most relevant component of the code is the particle sampler which, for a given hydrodynamic output on a particlization hypersurface, generates as many hadronic events (with full position and momentum information for all hadrons created in the event) as desired by the user.\footnote{%
    Some routines in the particle sampler algorithm were inspired by earlier work reported in \cite{Bernhard:2018hnz, Pang:2018zzo}.}
The particle sampler is validated by conducting high-precision tests comparing the event-averaged sampled momentum spectra and space-time distributions with the numerically evaluated continuous Cooper-Frye formula. The sampler output is compatible with and can be directly used as input for hadronic rescattering and kinetic freeze-out algorithms such as \U\ \cite{Bass:1998ca} and the recently developed hadronic afterburner code {\sc SMASH} \cite{Weil:2016zrk}.

This paper is organized as follows: In Sec.~\ref{S2} we review the four types of $\delta f_n$ corrections that the code can evaluate. In Sec.~\ref{S3} we describe the setup for producing the test hypersurfaces used to validate the particle sampler against the Cooper-Frye formula. In Sec.~\ref{S4} we discuss the implementation of the continuous Cooper-Frye formula, as well as its acceleration either on a multi-core processor with OpenMP or on a graphics processing unit (GPU) with CUDA. In Sec.~\ref{S5} we discuss the sampling of hadrons from the Cooper-Frye formula for each of the four $\delta f_n$ methods. Finally, we rigorously test the performance of the particle sampler in Sec.~\ref{S6}.

\section{Viscous corrections to the hadronic phase-space distribution}
\label{S2}
 
In this section, we summarize the four options for $\delta f_n$ that are available in the code package. A more detailed derivation of these different $\delta f_n$ corrections can be found in Ref. \cite{feqmod}. 

\subsection{Linearized viscous corrections}
\label{S2.1}

The 14-moment approximation and Chapman-Enskog expansion are the two most popular methods for linearizing the distribution function around $f_{\eq,n}$. Their expansion coefficients are adjusted to exactly reproduce $J_B^\mu$ and $T^\munu$. However, both methods suffer from the problem that, even for moderately large values of shear stress, bulk viscous pressure and baryon diffusion current, where according to modern understanding \cite{Romatschke:2017ejr, Heinz:2019dbd, Jaiswal:2019cju} dissipative hydrodynamics still works well, $f_{\eq,n} + \delta f_n$ can become negative at higher momentum, formally invalidating the truncation of the expansion at linear order and the interpretation of $f_n(x,p)$ as a probability from which particle positions and momenta can be sampled. This is not necessarily a big problem as long as one evaluates the Cooper-Frye formula by numerical integration to obtain a continuous function for the momentum spectra (describing their ensemble average over a large set of real events with a finite number of particles in each event), because one can simply integrate blindly over the regions where the phase-space density becomes negative, hoping that they do not carry much weight (and rejecting the result only when they do). However, when instead using the Cooper-Frye integrand to create real collision events, with a finite number of real particles emitted from the particlization surface, the regions of negative $f_n(x,p)$ must be cut out, for example by multiplying the integrand in Eq.~(\ref{eq1}) by hand with a step function enforcing $|\delta f_n| \leq f_{\eq,n}$. This modification of the Cooper-Frye formula slightly violates energy-momentum and charge conservation in the particlization process, and the right hand sides of Eqs.~\eqref{eq:JmuTmunu} no longer reproduce the hydrodynamic net-baryon current and energy-momentum tensor on the l.h.s. exactly. These violations are usually small but should be monitored by the user.

\subsubsection{14-moment approximation}
\label{S2.1.1}

The 14-moment approximation is an expansion of the dissipative correction $\delta f_n$ in momentum moments of the distribution function, truncated at the hydrodynamic level, i.e. at terms involving $p^\mu$ and $p^\mu p^\nu$. For a multicomponent relativistic gas with baryon chemical potential, the 14-moment approximation for $\delta f_n$ takes the form \cite{Israel:1976tn, Israel:1979wp, Monnai:2009ad, Denicol:2012cn}
%
\begin{equation}
\label{eq14moment}
  \delta f_n = 
  f_{\eq,n} \bar{f}_{\eq,n} \left(b_n c_\mu p^\mu + c_{\mu\nu} p^{\mu}p^{\nu}\right), 
\end{equation}
where $\bar{f}_{\eq,n} \equiv 1 \,-\, g_n^{-1} \Theta_n f_{\eq,n}$. To simplify the calculation we assume that the irreducible components of the coefficients $c_{\mu}$ and $c_{\mu\nu}$ are species independent. An irreducible contraction of the tensors in Eq.~\eqref{eq14moment} leads to \cite{Monnai:2009ad} 
\begin{equation}
\label{eq:14}
\begin{split}
  \delta f_n = & f_{\eq,n} \bar{f}_{\eq,n} 
  \Big( c_T \, m_n^2 + b_n \big(c_B (u \cdot p) + c_{V}^{\langle\mu\rangle}p_{\langle\mu\rangle}\big) + c_E(u \cdot p)^2\\
      &  + c_{Q}^{\langle\mu\rangle} (u \cdot p)p_{\langle\mu\rangle} + c_\pi^{\langle\mu\nu\rangle}
         p_{\langle\mu} p_{\nu\rangle} \Big)\,.
\end{split}
\end{equation}
One then fixes the coefficients such that $\delta f_n$ satisfies the Landau matching conditions $\delta \ene = \delta n_B =0$ (i.e. it doesn't contribute to the energy and net baryon density) and reproduces the shear stress tensor, bulk viscous pressure, and baryon diffusion current:
\bs
\beal
   c_T &= A_T \Pi, \qquad
   c_B = A_B \Pi, \qquad
   c_E = A_E \Pi, \qquad \\
   c_{V}^{\langle\mu\rangle} &= A_V V_B^{\mu}, \qquad
   c_{Q}^{\langle\mu\rangle} = A_Q V_B^{\mu}, \qquad
   c_{\pi}^{\langle\mu\nu\rangle} = A_\pi \pi^{\mu\nu}.
\end{align}
\es
Here $A_T$, $A_B$, $A_E$, $A_V$, $A_W$ and $A_\pi$ are algebraic combinations of moments of the local equilibrium distribution $f_{\eq,n}$; their analytic expressions can be found in Appendix~\ref{App_A}.

\subsubsection{First-order Chapman-Enskog expansion}
\label{S2.1.2}

The Chapman-Enskog expansion is a gradient expansion around $f_{\eq,n}$ whose coefficients are derived from the Boltzmann equation \eqref{eq:boltzmann}. Here, we use the relaxation-time approximation (RTA \cite{Anderson_Witting_1974}) for the collision term $\mathcal{C}_n$ in which Eq.~(\ref{eq:boltzmann}) reduces to 
\begin{equation}
   p \cdot \partial f_n = 
   - \frac{u \cdot p}{\tau_r}\left(f_n{-}f_{\eq,n}\right)
   = - \frac{u \cdot p}{\tau_r} \delta f_n\,.
\end{equation}
The relaxation time $\tau_r$ is assumed to be momentum and species independent. Substituting the decomposition (\ref{eq-decomp}) on the l.h.s. and assuming that the hydrodynamic gradients are small compared to the relaxation rate $\tau^{-1}_r$, one can neglect the derivatives of $\delta f_n$ on the l.h.s. and derive a first-order gradient correction to the thermal distribution \cite{Jaiswal:2014isa}:\footnote{%
    This equation nicely illustrates the competition between global expansion (causing the gradients in the numerator) and local scattering (in the denominator) in establishing the size of the deviation $\delta f_n$ from local equilibrium.}  
\begin{equation}
\label{eqChapman}
\delta f_n = - \frac{p \cdot \partial f_{\eq,n}}{p \cdot u/\tau_r}\,.
\end{equation}
After expanding out the partial derivative one uses the conservation of net baryon number, energy and momentum in the form
\begin{equation}
\label{eq:Tdot}
  \dot{\alpha}_B = \mathcal{G} \theta, \qquad
  \dot{T} = \mathcal{F} \theta, \qquad
  \dot{u}^\mu = \Delta^{\mu\nu} \partial_\nu \ln{T}\,,
\end{equation}
where the dots denote the LRF time derivative, $\dot A = u^{\mu} \partial_{\mu} A$), together with the first-order Navier-Stokes relations
\begin{equation}
  \Pi = - \zeta \theta, \qquad
  \pi^{\mu\nu} = 2\eta \sigma^{\mu\nu}, \qquad
  V_B^{\mu} = \kappa_B \Delta^{\mu\nu} \partial_\nu  \alpha_B\,,
\end{equation}
where $\zeta$ and $\eta$ are the bulk and shear viscosity and $\kappa_B$ is the baryon diffusion coefficient, to rewrite Eq.~\eqref{eqChapman} as
\begin{equation}
\label{eq:CE}
\begin{split}
    \delta f_n = &f_{\eq,n} \bar{f}_{\eq,n} 
    \Bigg[ \frac{\Pi}{\beta_\Pi}
           \left(b_n \mathcal{G} + \frac{(u\cdot p)\mathcal{F}}{T^2} 
               + \frac{(-p\cdot\Delta\cdot p)}{3 (u \cdot p) T} 
           \right) \\
         & + \frac{V_B^\mu p_{\langle\mu\rangle}}{\beta_V} \left(\frac{n_B}{\ene + \Peq} - \frac{b_n}{(u \cdot p)}\right) + \frac{\pi_{\mu\nu} p^{\langle\mu}p^{\nu\rangle}}
                {2\beta_{\pi} (u\cdot p) T}
    \Bigg];
\end{split}
\end{equation}
here $\beta_{\pi} = \eta / \tau_r$, $\beta_{\Pi} = \zeta / \tau_r$, and $\beta_{V} = \kappa_B / \tau_r$ are the ratios of  the viscosities and baryon diffusion coefficient to their respective relaxation times. As for the 14-moment coefficients, the Chapman-Enskog coefficients are assumed to be species independent and are adjusted to reproduce the energy-momentum tensor of the fluid. The functions $\mathcal{G}$ and $\mathcal{F}$ in (\ref{eq:Tdot}) and the coefficients $\beta_\pi$, $\beta_\Pi$, and $\beta_V$ in (\ref{eq:CE}) are all given by thermal integrals over the sum of all equilibrium contributions $f_{\eq,n}$ (see Appendix~\ref{appa1}) and listed in Appendix~\ref{appa2}. 

\subsection{Modified equilibrium distribution}
\label{S2.2}

Modified equilibrium distributions address the negative probability problem in the linearized approaches by assuming the same functional form as $f_{\eq,n}$ but with rescaled effective temperature and particle momenta that depend on the shear stress, bulk pressure and baryon diffusion current. In this way, the distribution function remains positive-definite while also reproducing the energy-momentum tensor to a good degree of accuracy. 

\subsubsection{The Pratt-Torrieri-Bernhard {\rm (PTB)} distribution}
\label{S2.2.1}

Bernhard's distribution \cite{Bernhard:2018hnz} is a variant of the modified equilibrium distribution initially proposed by Pratt and Torrieri \cite{Pratt:2010jt}. This {\it Pratt-Torrieri-Bernhard} (PTB) distribution is defined by
\begin{equation}
\label{jonah}
  f_{\eq,n}^{\text{PTB}} = \frac{\mathcal{Z}_\Pi}{\det A} \, g_n 
  \left[\exp\left( \frac{\sqrt{p'_i \, p'_i + m_n^2}}{T} \right) + \Theta_n\right]^{-1},
\end{equation}
where $\mathcal{Z}_\Pi>0$ is a positive renormalization factor and $A_{ij}$ is the momentum transformation matrix, both specified further below.\footnote{%
    The baryon chemical potential and diffusion current are set to zero in the PTB distribution \cite{Bernhard:2018hnz}.} 
The distribution function \eqref{jonah} is isotropic in the momentum $\bm{p}'$. The components of $\bm{p}'$ are related to the spatial components of the particle momentum $p^\mu$ in the LRF, $p_i = -X_i \cdot p$,\footnote{\label{fn10}%
    Here $X_i^\mu = (X^\mu, Y^\mu, Z^\mu)$ are the spatial basis vectors pointing along the $x,y$ and $z$ directions in the LRF, i.e.\ $X^\mu_{_{\LRF}} = (0,1,0,0)$, $Y^\mu_{_{\LRF}}=(0,0,1,0)$, and $Z^\mu_{_{\LRF}}=(0,0,0,1)$.}
by the linear transformation
\begin{equation}
\label{rescaling}
  p_i = A_{ij}p'_j\,.
\end{equation}
The matrix $A_{ij}$ encodes momentum space deformations caused by the viscous pressures:
\begin{equation}
\label{Lambda}
  A_{ij} = (1+\lambda_{\Pi})\delta_{ij}  
           + \frac{\pi_{ij}}{2\beta_\pi}\,.
\end{equation}
Here $\pi_{ij} \equiv X_i{\,\cdot\,}\pi{\,\cdot\,}X_j$ are the spatial LRF components of the shear stress tensor $\pi^\munu$. In the absence of shear stress, the parameters $\lambda_{\Pi}$ and $\mathcal{Z}_{\Pi}$ are adjusted to reproduce the total pressure $\mathcal{P}_\eq + \Pi$ without changing the energy density. For zero bulk viscous pressure, the shear term in the momentum deformation matrix (\ref{Lambda}) is constructed such that for small values of $\pi^{\mu\nu}$ one recovers the shear correction of the Chapman-Enskog expansion \eqref{eq:CE}. These adjustments are not further modified when both the shear and bulk viscous pressures are non-zero \cite{Bernhard:2018hnz}.

After inverting the linear transformation \eqref{rescaling} and inserting the result into (\ref{jonah}) the resulting modified equilibrium distribution is positive definite as long as $\det A{\,>\,}0$. Since the factor $\mathcal{Z}_\Pi/\det{A}$ is the same for all hadron species, the viscous pressures do not alter the particle abundance ratios from their equilibrium values.\footnote{%
    In general, this renormalization factor does not conserve the net baryon number, restricting the application of the PTB distribution to cases where either the net baryon current $J_B^{\mu}$ or bulk viscous pressure $\Pi$ vanishes.}

\subsubsection{The {\rm PTM} distribution}
\label{S2.2.2}

The PTM distribution, which improves upon the original modified equilibrium distribution of Pratt and Torrieri \cite{Pratt:2010jt}, is defined by the formula~\cite{feqmod}
\begin{equation}
\label{Mike}
    f_{\eq,n}^{\text{PTM}} = \mathcal{Z}_n \, g_n 
    \left[\exp\left( \frac{\sqrt{p'_i \, p'_i + m_n^2}}
                        {T + \beta^{-1}_{\Pi}\Pi\mathcal{F}} - b_n\left(\alpha_B +  \beta^{-1}_{\Pi}\Pi\mathcal{G}\right)
            \right) + \Theta_n
    \right]^{-1}\,.
\end{equation}
It differs from the PTB distribution (\ref{jonah}) by different choices for the renormalization factor (which is now species dependent) and the momentum-transformation. Furthermore, a non-zero bulk viscous pressure now modifies the effective temperature and baryon chemical potential of the modified equilibrium distribution. For PTM the momentum transformation 
(\ref{rescaling}) is
\begin{equation}
\label{eq:transform_Mike}
  p_i = A_{ij} p'_j \,-\, q_i \sqrt{{p'}^2 + m_n^2} + b_n T a_i
\end{equation}
where 
\bs
\beal
\label{eq:matrix_Mike}
    A_{ij} &= \left(1 + \frac{\Pi}{3\beta_\Pi}\right) \delta_{ij} 
           + \frac{\pi_{ij}}{2\beta_\pi}\,, \\
\label{eq:baryon_Mike}
    q_{i} &= \frac{V_{B,i} \, n_B T}{\beta_V (\ene + \Peq)}\,, \\
    a_{i} &= \frac{V_{B,i}}{\beta_V}\,.
\end{align}
\es
The shear stress term in the matrix $A_{ij}$ is the same as in Eq.~\eqref{Lambda} but the bulk viscous pressure $\Pi$ now rescales the momenta differently \cite{Pratt:2010jt}. Additionally, the momenta are shifted along the direction of the LRF baryon diffusion current $V_{B,i} = - X_i \cdot V_B$. The renormalization factor $\mathcal{Z}_n$ in  Eq.~\eqref{Mike} is now adjusted to reproduce the linearized particle density given by the Chapman-Enskog expansion; it is defined as
$\mathcal{Z}_n = n^{(1)}_n / n^{\mathrm{(raw)}}_n$ where
%
\bs
\beal
\label{linear_density}
    n^{(1)}_n &= n_{\eq,n}(T,\alpha_B)
              + \frac{\Pi}{\beta_{\Pi}}\left(n_{\eq,n}(T,\alpha_B) 
              + \mathcal{N}_{10}^{(n)} \mathcal{G}
              + \frac{\mathcal{J}_{20}^{(n)}\mathcal{F}}{T^2}\right) \\
\ \ 
    n^{\text{(raw)}}_n &= \det{A} \times n_{\eq,n}
    \left(T + \beta^{-1}_{\Pi}\Pi\mathcal{F}, \alpha_B + \beta^{-1}_{\Pi}\Pi\mathcal{G}\right)
\end{align}
\es
are the linearized and the raw (i.e. un-renormalized) PTM particle densities, respectively, with $n_{\eq,n}(T,\alpha_B)$ being the equilibrium particle density. The functions $\mathcal{N}_{10}^{(n)}$ amd $\mathcal{J}_{20}^{(n)}$ is listed in App.~\ref{App_A}. For nonzero bulk viscous pressure the renormalization factor $\mathcal{Z}_n$ becomes species dependent and thus affects the particle abundance ratios. In the limit of small viscous pressures and diffusion current the PTM distribution reduces to the Chapman-Enskog expansion~\eqref{eq:CE}. 

\subsubsection{Breakdown of the modified equilibrium distributions}
\label{S2.2.3}

The modified equilibrium distributions defined in the preceding subsections have the advantage over the linearized forms described in Sec.~\ref{S2.1} that they are positive definite and can directly be used for sampling, without further intervention.
They are applicable for moderate viscous corrections but for large dissipative flows it is possible to encounter either a negative Jacobian determinant
\be
\label{eq:Jacobian}
\det\left(\frac{\partial p_i}{\partial p^\prime_j}\right) = \det{A} \left(1 - \frac{q_i A^{-1}_{ij}p^\prime_j}{\sqrt{(p^{\,\prime})^2 + m_n^2}}\right)
\ee
(where $q_i = 0$ in the PTB case) or a negative renormalization factor $\mathcal{Z}_n$ (this only happens in the PTM case). This is not supposed to happen as long as the viscous hydrodynamic code operates within its range of applicability, but in practice it can nevertheless occur in limited regions of the particlization surface, for a number of technical reasons. When it happens, the modified equilibrium distribution fails to reproduce the energy-momentum tensor and net baryon current accurately, and we switch back to a linearized version of the modified distribution.\footnote{%
    This is not to say that in such regions the linearized form
    for $\delta f_n$ represents a physically acceptable approximation; it is only a technical fix to limit the deviations of the energy-momentum tensor and net-baryon current on the particlization hypersurface.}
For the PTM distribution, this linearized $\delta f_n$ correction coincides with the first-order Chapman-Enskog expansion (\ref{eq:CE}). The linearized version of the PTB distribution is
\be
\label{eq:jonah_linear}
    \delta f_n \approx f_{\eq,n} 
    \left[\delta\mathcal{Z}_\Pi - 3 \delta\lambda_\Pi 
          +\bar{f}_{\eq,n}
          \left(\frac{\delta\lambda_\Pi(- p\cdot\Delta\cdot p)}
                     {(u\cdot p)T} 
               +\frac{\pi_{\mu\nu} p^{\langle\mu}p^{\nu\rangle}}
                     {2\beta_{\pi} (u \cdot p) T}
          \right) 
    \right]\,,
\ee
where $1+\delta\mathcal{Z}_\Pi$ and $\delta\lambda_\Pi$ are the renormalization and isotropic momentum scale factors evaluated in the limit of small bulk viscous pressure (see Ref.~\cite{feqmod} for more details).\footnote{%
    In the Bayesian analysis presented in \cite{Bernhard:2018hnz}, Bernhard always samples hadrons using the PTB distribution and never switches to the linearized $\delta f_n$ correction \eqref{eq:jonah_linear} -- not even when the former distribution breaks down.}
The coefficients $(\delta \mathcal{Z}_\Pi, \delta\lambda_\Pi)$ are given in Appendix~\ref{appa3}.

\section{Setup}
\label{S3}

The goal of this work is to describe and validate the \I\ particle sampler, by testing the event-averaged sampled particle spectra and space-time distributions against the continuous results obtained directly from the Cooper-Frye integrals, computed for all four $\delta f_n$ corrections described in the preceding section. In this section, we describe the collision systems used to generate the particlization hypersurfaces used in our tests.

\subsection{Central Pb-Pb collision}
\label{S3.1}

We use the code GPU-VH \cite{Bazow:2016yra} to evolve a longitudinally boost-invariant central Pb-Pb collision with (2+1)-dimensional viscous hydrodynamics.\footnote{%
    For the purpose of testing and validating the \I\ sampler the assumption of a boost-invariant hydrodynamic medium is not critical. The sampler code itself does not make use of this symmetry.}     
The GPU-VH code uses a fixed (Eulerian) computational grid in Milne coordinates $x^\mu=(\tau,\bm{x}_\perp,\eta_s)$ where $\bm{x}_\perp=(x,y)$ are Cartesian coordinates in the plane transverse to the beam direction $z$, $\tau=\sqrt{t^2{-}z^2}$ is the longitudinal proper time, and $\eta_s = \frac{1}{2} \ln[(t{+}z)/(t{-}z)]$ is the space-time rapidity along the beam direction. For longitudinally boost-invariant systems the distribution function $f_n(x,p)= f_n(\tau,\bm{x}_\perp, \bm{p}_\perp, \eta_s{-}y_p)$ can depend only on the difference between the space-time rapidity $\eta_s$ and momentum rapidity $y_p=\frac{1}{2}\ln[(E{+}p_z)/(E{-}p_z)]$, and macroscopic fields (e.g. the temperature $T(x)$) must be independent of $\eta_s$.   

Due to the assumed longitudinal boost-invariance we are only interested in midrapidity observables at $y_p{\,=\,}\eta_s{\,=\,}0$. We use smooth optical Glauber initial conditions \cite{Miller:2007ri} with a central temperature of $T_0 = 600$ MeV. We start the hydrodynamic simulation at a longitudinal proper time $\tau_0 = 0.25$ fm/$c$, with the spatial components of the fluid velocity $u^i$, shear stress tensor $\pi^\munu$ and bulk pressure $\Pi$ initialized to zero. We set the specific shear viscosity to $\eta / \mathcal{S} = 0.2$. The specific bulk viscosity is parameterized as $(\zeta / \mathcal{S})(T) = (\zeta/\mathcal{S})_\mathrm{norm} \, f(T/T_p)$, where 
%
\begin{figure*}[t!]
\centering
\includegraphics[width=\textwidth]{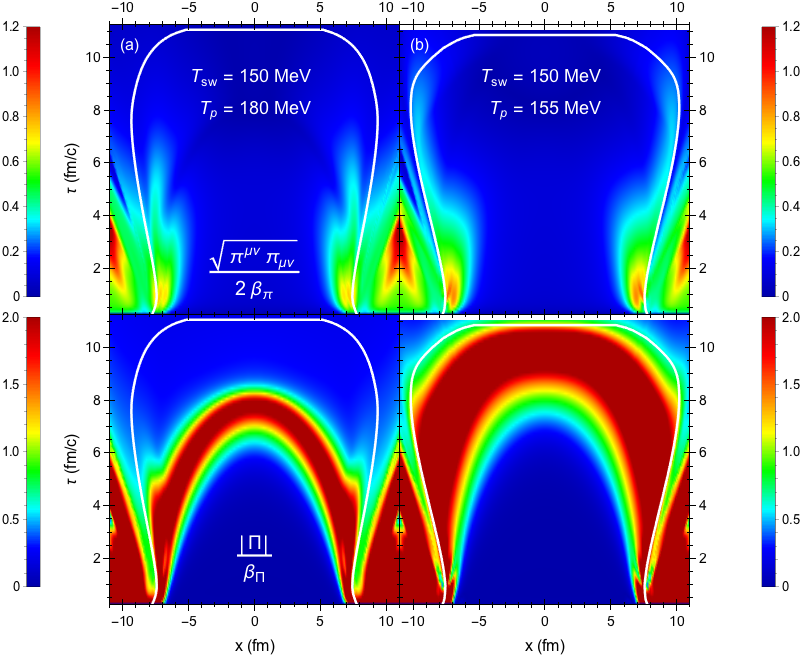}
\caption{(Color online)
    The $(\tau, x)$ slice at $y{\,=\,}\eta_s{\,=\,}0$ of the shear Knudsen number $\mathrm{Kn}_\pi{\,=\,}\sqrt{\pi^\munu \pi_\munu} / (2\beta_\pi)$ (top panels), bulk Knudsen number $\text{Kn}_{_\Pi} = |\Pi| / (\beta_\Pi)$ (bottom panels), and particlization hypersurface at temperature $T_\mathrm{sw} = 150$\,MeV (white contour) from the hydrodynamic simulation of a (2+1)-d central Pb-Pb collision with smooth Glauber initial conditions. Results for two choices for the temperature $T_p$ at which the specific bulk viscosity $\zeta / \mathcal{S}$ peaks ($T_p=180$\,MeV (a) and $T_p=155$\,MeV (b)) are shown in the left and right columns. 
\label{FHydro}
\vspace*{-5mm}
}
\end{figure*}
%
%
\be
f(x) = \Bigg\{
\begin{array}{ll}
      C_1 + \lambda_1 \exp\big[\frac{x-1}{\sigma_1}\big] + \lambda_2\exp\big[\frac{x-1}{\sigma_2}\big] 
      & (x{\,<\,}0.995), \\
      A_0 + A_1 x + A_2 x^2 & (0.995{\,\leq\,}x{\,\leq\,}1.05), \\
      C_2 + \lambda_3 \exp\big[\frac{1-x}{\sigma_3}\big] + \lambda_4\exp\big[\frac{1-x}{\sigma_4}\big]  & (x{\,>\,}1.05),
\end{array} 
\ee
with parameters $A_0 = -13.45$, $A_1 = 27.55$, $A_2 = -13.77$, $C_1 = 0.03$, $C_2 = 0.001$, $\lambda_1 = 0.9$, $\lambda_2 = 0.22$, $\lambda_3 = 0.9$, $\lambda_4 = 0.25$, $\sigma_1 = 0.0025$, $\sigma_2 = 0.022$, $\sigma_3 = 0.025$ and $\sigma_4 = 0.13$ \cite{Denicol:2009am}. We set the normalization factor to ($\zeta/\mathcal{S})_\text{norm} = 1$ and the peak temperature to either $T_p{\,=\,}180$\,MeV or 155\,MeV, as specified in each case below.\footnote{%
    Note that the (3+1)-d hydrodynamic code GPU-VH \cite{Bazow:2016yra} does not evolve the net baryon density and baryon diffusion current. In this work we therefore set $\alpha_B = 0 = V_B^\mu$. While the baryon sector is fully implemented in \I\ it has so far not been tested. We plan to provide the corresponding tests in the near future using hydrodynamic output from the recently developed (3+1)-d code {\sc BEShydro} \cite{Du:2019obx}.}

A particlization hypersurface of constant temperature $T_\mathrm{sw}{\,=\,}150$ MeV is generated using the freezeout surface finder code CORNELIUS \cite{Huovinen:2012is}. For a central Pb-Pb collision with these parameters, the boost-invariant hypersurface contains about $N_\Sigma = 1.9 \times 10^5$ freezeout cells, with a temporal and spatial resolution of approximately $\Delta\tau \approx 0.05$\,fm/$c$ and $(\Delta x, \Delta y) \approx 0.1$\,fm, respectively.\footnote{%
    These cells are located at space-time rapidity $\eta_s{\,=\,}y_{p}^\text{CM}$ where $y_{p}^\text{CM}$ is the center-of-momentum rapidity of the collision system ($y_{p}^\text{CM}{\,=\,}0$ in this work). The Cooper-Frye integral involves an integral over $\eta_s$; in this work the freeze-out information at space-time rapidities $\eta_s{\,\ne\,}y_{p}^\text{CM}$ is generated analytically from the freeze-out information at $\eta_s{\,=\,}y_{p}^\text{CM}$ using boost invariance. For hydrodynamic output from a genuinely (3+1)-d simulation (i.e. starting from initial conditions without longitudinal boost-invariance) this analytically generated input must be replaced by hydrodynamic output at different space-time rapidities $\eta_s{\,\ne\,}y_{p}^\text{CM}$, increasing the number of freeze-out cells accordingly.}
Fig.~\ref{FHydro} shows the $(\tau,x)$ slice at $y{\,=\,}\eta_s{\,=\,}0$ of the shear and bulk Knudsen numbers as well as the particlization surface generated from the simulation.

For central collision systems we are interested in the azimuthally averaged transverse momentum spectra
\be
\label{eq:central_pT_spectra}
    \frac{dN_n}{2\pi p_T dp_T dy_p} = \int_0^{2\pi} 
    \frac{d\phi_p}{2\pi} \frac{dN_n}{p_T dp_T d\phi_p dy_p}
    = \int_0^{2\pi} \frac{d\phi_p}{2\pi}
      \frac{1}{(2\pi\hbar)^3}\int_\Sigma p \cdot d^3\sigma \, f_n
\ee
as well as the temporal and (azimuthally averaged) radial distributions 
\begin{eqnarray}
\label{eq:time_formula}
    \frac{dN_n}{\tau d\tau d\eta_s} &=& 
    \frac{\partial^2}{\tau \partial\tau \partial\eta_s}
    \int_p \int_\Sigma p \cdot d^3 \sigma \, f_n, 
\\
\label{eq:radial_formula}
    \frac{dN_n}{2\pi r dr d\eta_s} &=& 
    \frac{\partial^2}{2\pi r \partial r \partial\eta_s}
    \int_p \int_\Sigma p \cdot d^3 \sigma \, f_n \,,
\end{eqnarray}
where $r{\,=\,}\sqrt{x^2{+}y^2}$. Due to the azimuthal symmetry of the optical Glauber model initial condition (which represents an ensemble average over fluctuating initial conditions with random orientations in the transverse plane) there are no interesting azimuthally sensitive observables to be computed from this particlization surface.\footnote{%
    Azimuthal fluctuations arising from finite-number statistical effects in the individually sampled events are of physical interest but without value for code verification. They will be part of a separate study of hydrodynamic model predictions using the \I\ sampler.}

\subsection{Non-central Pb-Pb collision}
\label{S3.2}

As an example for a non-central collision fireball we evolve, for the same Glauber model and viscosity parameters, a smooth hydrodynamic event with nonzero impact parameter $b{\,=\,}5$\,fm. The resulting particlization surface emits particles with anisotropic flow which is encoded in the differential flow coefficients
\be
\label{eq:vn}
    v_k^{(n)}(p_T) = 
    \frac{\displaystyle{\int_0^{2\pi}} \!\! d\phi_p\,    
          e^{ik\phi_p}\,\frac{dN_n}{p_T\,dp_T\,d\phi_p dy_p}}
         {\displaystyle{\int_0^{2\pi}} \!\! d\phi_p\,  
         \frac{dN_n}{p_Tdp_Td\phi_p dy_p}}
    = \frac{\displaystyle{\int_0^{2\pi}} \!\! d\phi_p\,    
          e^{ik\phi_p}\,\! \int_\Sigma p \cdot d^3\sigma\,f_n}
         {\displaystyle{\int_0^{2\pi}} \!\! d\phi_p\,  
         \! \int_\Sigma p \cdot d^3\sigma\,f_n}.
\ee
In particular, we will be interested in computing the elliptic and quadrangular flow coefficients $v_2(p_T)$ and $v_4(p_T)$ for non-central collisions.\footnote{%
    Due to the $x\leftrightarrow-x$ reflection symmetry of the fireball in the optical Glauber limit all odd flow coefficients vanish, and the factor $e^{ik\phi_p}$ under the integral in (\ref{eq:vn}) can be replaced by $\cos(k\phi_p)$.}
    
\section{Continuous Cooper-Frye distributions}
\label{S4}

The Cooper-Frye formulae (\ref{eq:central_pT_spectra}-\ref{eq:vn}) describe continuous distributions that can be interpreted as the statistical ensemble average of the fluctuating distributions for individual collision events obtained by interpreting the Cooper-Frye integrand $p\cdot d^3\sigma\,f_n$ as a probability distribution. We will use them to check the accuracy and performance of the \I\ sampler. In this section we review the numerical computation of Eqs.~(\ref{eq:central_pT_spectra}-\ref{eq:vn}) for longitudinally boost-invariant (2+1)-d and general (3+1)-d hypersurfaces. To give the user the option to speed up the calculation on either a multi-core CPU or a GPU, we provide in the code package two versions of the same algorithm, one in C++ with OpenMP and the other in CUDA.\footnote{%
    Note that only the Cooper-Frye integration for continuous spectra is parallelized but not the \I\ particle sampler.} 

\subsection{Integration routine}
\label{S4.1}
To compute the continuous transverse momentum spectra \eqref{eq:central_pT_spectra} and aniso\-tropic flow coefficients \eqref{eq:vn} we integrate the Cooper-Frye formula numerically:
\be
\label{eq:continuous_spectra}
    \frac{dN_n}{p_T dp_T d\phi_p dy_p} = 
    \sum^{N_\Sigma}_i p \cdot d^3\sigma_i \, 
    f_n(x^\mu_i, p_T,\phi_p, y_p)\,.
\ee
Here $d^3\sigma_i$ are the discrete hypersurface elements at positions $(\tau_i, x_i, y_i, \eta_{s,i})$. The algorithm for computing the momentum spectra \eqref{eq:continuous_spectra} is straightforward: one simply loops over the freezeout cells $i$ and adds their contribution to the spectra of each particle species at different momentum points. After integrating over the freezeout surface, we use Gaussian quadrature to compute the observables~\eqref{eq:central_pT_spectra} and~\eqref{eq:vn}.

An exact calculation of the space-time distributions \eqref{eq:time_formula} and \eqref{eq:radial_formula} requires knowledge about the partial derivatives of $d^3 \sigma$ and $f_n$. Since the freezeout surface finder code does not provide this information, we use a zeroth-order approximation by computing the particle yield from each freezeout cell
\be
\label{eq:yield_per_cell}
  \Delta N_{n,i} = \int p_T \, dp_T \, d\phi_p \, dy_p \, p \cdot d^3\sigma_i \, f_n(x^\mu_i, p_T,\phi_p, y_p)\,,
\ee
which is evaluated with Gaussian quadrature. We then construct the space-time distributions by binning the weights \eqref{eq:yield_per_cell} in a uniform space-time grid with $\Delta \tau = 0.1$ fm/c, $\Delta r = 0.2$ fm, and $\Delta \eta_s = 0.1$.

For the longitudinally boost-invariant test hypersurfaces used in this work the integration routine is slightly modified. The momentum spectra are evaluated using the formula
\be
\label{eq:continuous_spectra_2D}
  \frac{dN_n}{p_T dp_T d\phi_p dy_p} = 
  \sum^{N_\Sigma}_i   p{\cdot}d^2\sigma_i 
  \sum^{N_{\eta_s}}_j \omega_j\, f_n
  \bigl(x^\mu_{\perp i}, p_T,\phi_p,(y_p - \eta_s)_j\bigr).
\ee
Here the first sum over $i$ contains only hydrodynamically generated freeze-out cells in the transverse plane at space-time rapidity $\eta_s{\,=\,}y_{p}^\text{CM}$, i.e. (2+1)-dimensional surface elements $d^2\sigma_i$ at positions $x^\mu_{\perp i} = (\tau_i, x_i, y_i, \eta_s{=}y_{p}^\text{CM})$. The second sum over $j$ represents the integration over $\eta_s$, using a grid centered around the rapidity $y_{p}^\text{CM}$ and integration weights $\omega_j$. The default 48-point grid $(y_{p}^\text{CM}{-}\eta_s)_j$ and integration weights $\omega_j$ used in \I\ are
\bs
\label{eq:rapidity_grid}
\beal
  (y_{p}^\text{CM}{-}\eta_s)_j &= \sinh^{-1}\left(\frac{x_j}{1-x_j^2}\right) \\
\omega_j &= w_j \, \frac{1+x_j^2}{\big|1{-}x_j^2\big|\,\sqrt{1-x_j^2+x_j^4}}
\end{align}
\es
where $x_j$ and $w_j$ are the Gauss-Legendre roots and weights, respectively. If $f_n(x,p)$ is a modified equilibrium distribution, we follow Ref.~\cite{feqmod} and use an adaptive grid that adjusts itself from one freezeout cell to another, defined by
\bs
\beal
  (y_{p}^\text{CM}{-}\eta_s)_{i,j} &= \frac{\det A_i}{(\det A_{\Pi,i})^{2/3}}\, \times\,(y_{p}^\text{CM}{-}\eta_s)_j \\
\omega_{i,j} &= \frac{\det A_i}{(\det A_{\Pi,i})^{2/3}}\,\times\,\omega_j .
\end{align}
\es
Here $\det A_\Pi$ is the determinant of the momentum transformation matrix, given in Eqs.~\eqref{Lambda} and \eqref{eq:matrix_Mike} for the PTB and PTM distributions, respectively, but without the shear stress deformation. Table~\ref{tab:momentum_table} summarizes the formulas used for the remaining momentum grids $(p_T, \phi_p)$. 
\begin{table}[t]
\centering
\setlength{\tabcolsep}{0.875em} 
{\renewcommand{\arraystretch}{2.25}
 \begin{tabular}{|c|c|c|} 
 \hline
 & $\dfrac{dN}{2\pi p_T dp_T dy_p}$ \,\,\,\,\,\,\,\, $v_k(p_T)$ & \,\,\,$\dfrac{dN}{\tau d\tau d\eta_s}$ \,\,\,\,\,\,\,\, $\dfrac{dN}{2\pi rdr d\eta_s}$\,\,\,\\ [0.875ex]
 \hline 
 $p_{T,j} \, (\mathrm{GeV})$ & $0.03j - 0.015\,\,\,$ & $\left(\dfrac{1+x_j}{1-x_j}\right)^{1/2}$ \\ [0.875ex]\hline
 $\phi_{p,j}$ & $\pi(1+x_j)$ & $\pi(1+x_j)$ \\ [0.875ex]\hline 
 $(y_{p}^\text{CM}{-}\eta_s)_j$ & $\sinh^{-1}\left(\dfrac{x_j}{1-x_j^2}\right)$ & $\sinh^{-1}\left(\dfrac{x_j}{1-x_j^2}\right)$ \\ [0.875ex]
 \hline
 \end{tabular}}
 \caption{The momentum tables used in the continuous Cooper-Frye algorithm to compute the momentum spectra and space-time distributions for a 2+1d hypersurface. We use a 100-point uniform $p_T$ grid for the momentum spectra. The remaining entries use a 48-point non-uniform grid, where $x_j$ are the roots of the Legendre polynomial $P_{48}(x)$.}
\label{tab:momentum_table}
\end{table}

The space-time distributions for a longitudinally boost-invariant (2+1)-d hypersurface are $\eta_s$-independent. In this case, we evaluate the particle yield per unit space-time rapidity of each freezeout cell
\be
\label{eq:dNdeta_per_cell}
  \frac{\Delta N_{n,i}}{\Delta\eta_s} = \int p_T \, dp_T \, d\phi_p \, dy_p \, p \cdot d^2\sigma_i \, f_n(x^\mu_{\perp i}, p_T,\phi_p,y_p{-}\eta_s)\,,
\ee
and bin the weights in the ($\tau,r$)-grid.\footnote{%
    For the rapidity integral, we use the same Gaussian quadrature~\eqref{eq:rapidity_grid} except the $y_p$ grid is centered around the space-time rapidity $\eta_s = y_{p}^\text{CM}$.}

Although both of these integration routines are relatively simple, the number of numerical evaluations required is staggering. Let us consider the example of computing the spectra of all the particles that can propagate in {\sc SMASH}, which include about $N_R=444$ different hadron resonance species. Even for a 2+1d hypersurface, the number of freezeout cells, multiplied by the number of particles and momentum points, gives a total of $N = N_{d\sigma}{\,\times\,}N_\eta{\,\times\,}N_R{\,\times\,}N_{p_T}{\,\times\,}N_{\phi_p} = (1.9 \times 10^5) \times 48 \times 444 \times (48{-}100) \times 48 \sim (9{\,\times\,}10^{12} - 2{\,\times\,}10^{13})$ numerical calculations for central collisions. As a result, it becomes very time-consuming to compute the smooth Cooper-Frye formula on a single-core central processing unit (CPU). The amount of time it takes to compute the space-time distribution and momentum spectrum per particle with a high degree of accuracy on a single-core Intel Xeon E5-2680 v4 CPU with the Intel compiler and $-03$ optimization is about 626\,s and 1510\,s, respectively. In the next two subsections we describe how to speed up the algorithm with either OpenMP or CUDA.

\subsection{OpenMP Acceleration}

\begin{figure*}[t]
  \begin{center}
  \includegraphics[width=0.8\textwidth]{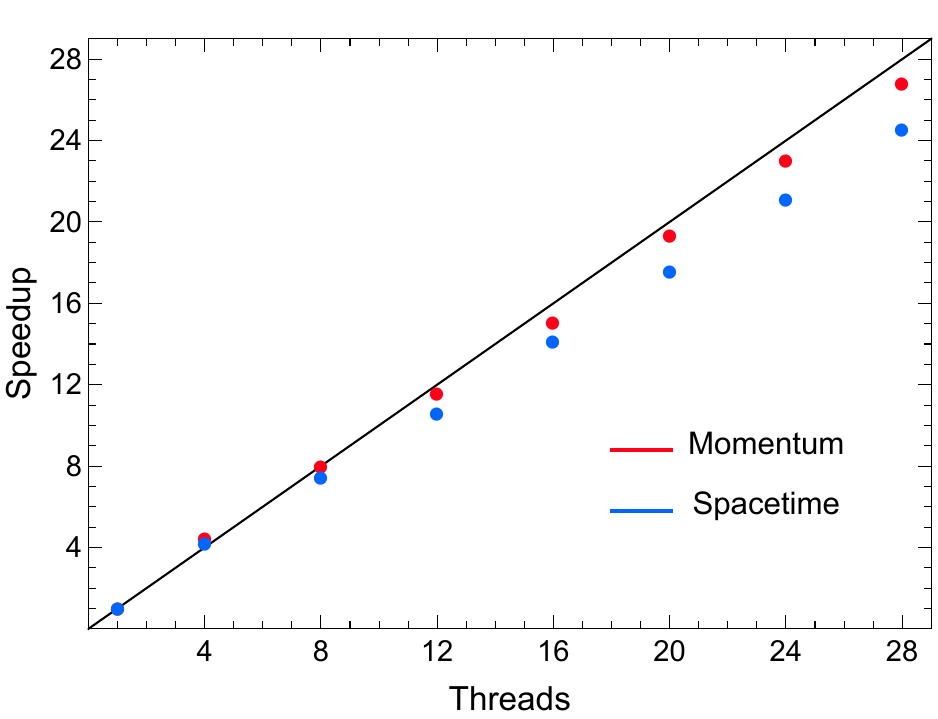}
  \end{center}
  \caption{(Color online)
    The speedup of the continuous particle momentum spectra (red dots) and space-time distribution (blue dots) routines on an Intel Xeon E5-2680 v4 multi-core processor with OpenMP. Ideally, the speedup scales with the number of CPU threads (solid black).
\label{F_OMP_speedup}
}
\end{figure*}

By default, the code is parallelized across multiple CPU threads using OpenMP. For our situation, each thread is given an equal fraction of the freezeout surface to integrate. Afterwards, the threads are summed together in a reduction step. Fig.~\ref{F_OMP_speedup} shows the speedup of the continuous Cooper-Frye algorithm with OpenMP. One observes that the speedup scales quite well with the number of threads. On an Intel Xeon E5-2680 v4 multi-core processor, which has 14 cores (28 threads in total), it now takes only about 25\,s and 56\,s to compute the space-time distribution and momentum spectrum per particle, respectively. Users wanting to execute the code in OpenMP compile the files stored in the {\sc cpp} folder of the code package referred to in footnote~\ref{fn7}.

\subsection{GPU Acceleration}
\subsubsection{Running the code with CUDA}

To accelerate the continuous Cooper-Frye formula on a GPU we parallelize the freezeout cells \cite{Pang:2018zzo}. Compared to a multi-core processor, however, a GPU contains thousands of cores, so we can integrate more freezeout cells simultaneously. 

The user can use GPU acceleration by compiling the files in the {\sc cuda} folder of the code package in footnote~\ref{fn7}. The CUDA integration routine is executed in two steps. First, we subdivide the freezeout surface into chunks, where the number of freezeout cells per chunk $N_\text{size}$ is a tunable parameter. We pass the surface chunks from the host to the device one at a time and launch a kernel that assigns a freezeout cell to each thread. The threads compute their contribution to the particle spectra or space-time distribution in parallel. We use a reduction algorithm to sum the threads in each block (we set the number of threads per block to 128); the blocks' contribution is then written to the device memory. Next, a separate reduction kernel is launched to sum the blocks together. After launching all the freezeout surface chunks on the GPU, the results are copied from the device to the host and written to disk. 

\begin{table}[t]
\centering
\setlength{\tabcolsep}{0.875em} 
{\renewcommand{\arraystretch}{1.5}
 \begin{tabular}{|c|c|c|} 
 \hline
 & Intel Xeon E5 & Nvidia Tesla V100 \\ [0.0ex]
 \hline 
Processor cores & 14 & 5120 \\ \hline
\makecell{Clock speed (GHz)} & 2.40 - 3.30 & 1.25 - 1.38 \\ \hline
\makecell{Memory bandwidth (GB/s)} & 76.8 & 900 \\ 
 \hline
 \end{tabular}}
 \caption{The technical specifications of the Intel Xeon E5-2680 v4 multi-core processor and Nvidia Tesla V100-PCIE graphics card.}
\label{tab:computer_table}
\end{table}
\subsubsection{GPU speedup benchmarks}

\begin{figure*}[b!]
\begin{center}
  \includegraphics[width=0.8\textwidth]{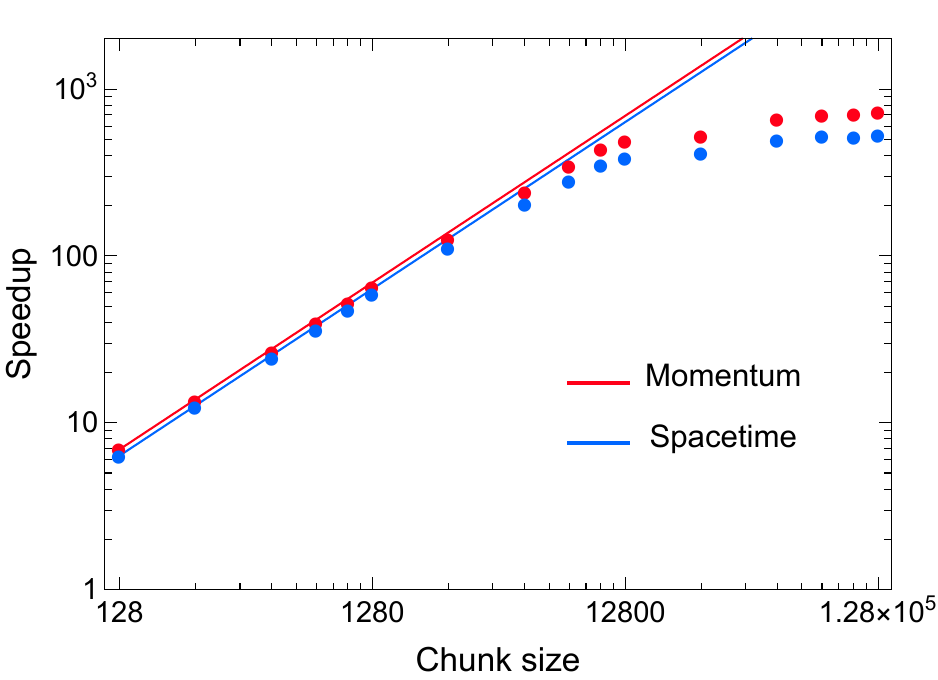}
\end{center}
  \caption{(Color online)
    The speedup of the continuous particle momentum spectra (red dot) and space-time distribution (blue dot) routines on an Nvidia Tesla V100-PCIE GPU as a function of the surface chunk size. A perfect parallelization across the freezeout cells gives a speedup proportional to the chunk size (solid color).
\label{F_GPU_speedup}
}
\end{figure*}

In this test we accelerate the evaluation of the continuous Cooper-Frye formula using the state-of-the-art Nvidia Tesla V100-PCIE graphics card; the technical specifications of the hardware is summarized in Table~\ref{tab:computer_table}. Fig.~\ref{F_GPU_speedup} shows the speedup of the  momentum and space-time integration routines as a function of the surface chunk size. One observes a speedup that scales with the surface chunk size because more freezeout cells are launched on the GPU simultaneously. However, the speedup begins to plataeu when the chunk size becomes so large that the threads cannot all be launched simultaneously. The main limitation is the number of cores on the GPU. The memory latency from writing the blocks' Cooper-Frye contributions to the device is also a limiting factor. Nevertheless, the amount of time it takes to compute the continuous Cooper-Frye formula is significantly reduced by a factor of about $500-700$.

\section{Sampling particles from the Cooper-Frye Formula} 
\label{S5}
 
The Cooper-Frye formula converts hydrodynamic output into hadron momentum spectra, but this conversion must be done before hydrodynamics breaks down. The final kinetic stage, in which the hadrons and hadronic resonances continue to rescatter (albeit at ever-decreasing rates) until they ultimately decouple and decay or free-stream to the detector, must be handled microscopically. This is usually done with the help of Monte Carlo implementations of kinetic equations in which real hadrons propagate on classical trajectories and rescatter stochastically. 

To initiate such a hadronic rescattering cascade requires the conversion of the hydrodynamic output on the switching surface into particles with positions and momenta, by interpreting the Cooper-Frye integrand as a probability density in phase-space and sampling it stochastically. Both theoretical arguments and model-to-data comparisons suggest that hydrodynamics is the more precise dynamical description of the fireball for temperatures above the pseudocritical temperature $T_c \approx 155$ MeV \cite{Borsanyi:2010cj, Borsanyi:2013bia, Bazavov:2014pvz}, whereas hadronic transport is a more reliable model below $T_c$ where the quark-gluon plasma liquid has fragmented into a gas of hadronic resonances \cite{Bernhard:2016tnd,Shen:2014lye}. In this section we describe the sampling mode of \I\ which provides such a \textit{particlization scheme}.  

One must keep in mind, however, that the integrand of Eq.~\eqref{eq1} is not always positive-definite, and this must be fixed before it can be used as a probability density. There are two possible sources of negative contributions: first, the particlization surface typically contains regions with spacelike normal vectors such that for certain ranges of momenta $p\cdot d^3\sigma < 0$, corresponding to particles being reabsorbed by the fireball. Second, when using a linearized form for the viscous correction to the distribution function $f_n$, the latter can turn negative at high momentum. To sample the Cooper-Frye integrand probabilistically it must be rendered positive-definite with the following modification:
\be
\label{eq:sampledCFF}
  E_p \frac{dN_n}{d^3p} = \frac{1}{(2\pi\hbar)^3}\int_\Sigma p \cdot d^3\sigma \, f_n \, \Theta(p \cdot d^3\sigma) \, \Theta(f_n) \,.
\ee
Here $\Theta$ is the Heaviside step function. As a consequence, the sampled particle spectra will deviate from the original Cooper Frye formula~\eqref{eq1} (e.g. energy-momentum and charge conservation are slightly violated). As will be discussed below, the deviations from the original spectra are typically small, except for very soft or hard momentum particles.

In the following sections, we discuss the methodology for sampling particles from Eq.~\eqref{eq:sampledCFF}, which is carried out in two steps: (i) For each freezeout cell we sample the number of hadrons emitted and their type. (ii) For each hadron produced, we then sample its momentum ~\cite{Pratt:2010jt, Bernhard:2018hnz, Shen:2014vra, Pang:2018zzo}.

\subsection{Sampling the number of hadrons from each freezeout cell}

Interpreting the number of particles given by the hydrodynamic output as the mean of a Poisson distribution, one can sample the number of hadrons $N$ from
\be
\label{eq:poisson}
  P(N) = \frac{\exp(-\Delta N_\text{h})(\Delta N_\text{h})^N}
              {N!} \,,
\ee
where
\be
\label{eq:meanHadrons}
\Delta N_\text{h}(x) = \sum_n \Delta N_n = \sum_n \int_p \, p \cdot d^3\sigma \, f_n \, \Theta(p \cdot d^3\sigma) \, \Theta(f_n)
\ee
is the mean number of hadrons emitted from the selected freezeout cell. After sampling the total number of hadrons, we sample their species from the discrete probability distribution 
\be
\label{eq:discrete}
D_n = \frac{\Delta N_n}{\Delta N_\text{h}} \,.
\ee
The C++ 11 library contains Poisson and discrete distribution classes which we use to sample Eqs.~\eqref{eq:poisson} and~\eqref{eq:discrete}, respectively. 

The prerequisite for sampling the numbers of hadrons is computing the mean number of each hadron species. However, enforcing the outflow of particles ($p \cdot d^3\sigma > 0$) in Eq.~\eqref{eq:meanHadrons} presents a complication in evaluating the momentum-space integral. If one ignores the effect of the function $\Theta(p \cdot d^3\sigma)$, then $\Delta N_n$ in Eq.~\eqref{eq:meanHadrons} reduces to (in the absence of diffusion current)
\be
\label{eq:meanHadronsNoOutflow}
\Delta N_n(x) \approx (u \cdot d^3\sigma) \int_p (u \cdot p) \, f_n \,\Theta(f_n) = \Bigl(u(x) \cdot d^3\sigma(x)\Bigr) \, n_n(x) \,,
\ee
which is simply the particle number density multiplied by the hypersurface volume element in the local rest frame. For timelike and lightlike freezeout cells the dot product $p \cdot d^3\sigma$ is always positive and Eq.~\eqref{eq:meanHadrons} reduces to Eq.~\eqref{eq:meanHadronsNoOutflow}. For spacelike cells, there are momentum space regions with $p \cdot d^3\sigma < 0$ that are cut out by the function $\Theta(p \cdot d^3\sigma)$, increasing the particle yield. If a considerable fraction of the emission occurs from spacelike domains on the hypersurface one must include the outflow effect on the mean hadron number.

Enforcing positivity of the distribution function adds another dimension of complexity to the evaluation of Eq.~\eqref{eq:meanHadrons}. If $f_n$ contains linear viscous corrections, then the function $\Theta(f_n)$ effectively regulates $\delta f_n$ such that 
\be
\label{eq:df_reg}
  \delta f_n \rightarrow \delta f_{n,\mathrm{reg}} = 
  \max\left(-f_{\eq,n},\,\min\left(\delta f_n, \,f_{\eq,n}\right)
      \right).
\ee
Here, we place an additional bound such that $|\delta f_{n,\text{reg}}| \leq f_{\eq,n}$ even if $\delta f_n$ is positive.\footnote{%
    Although the upper bound is not required, it facilitates the calculation of the maximum hadron number \eqref{eq:meanHadronsMax}. 
    } 
The primary culprit for causing negative $f_n$ is the linearized bulk viscous correction, causing it to be regulated to zero at high momentum. 

An exact evaluation of the mean hadron number \eqref{eq:meanHadrons} would require identifying the boundary along which the argument of one or the other $\Theta$ function vanishes, which is hard. We circumvent this problem with a stochastic trick \cite{Pratt:2010jt,Bernhard:2018hnz}, by making use of the inequality
\be
  \int_p p \cdot d^3\sigma \,\Theta(p \cdot d^3\sigma) \, 
  \left(f_{\eq,n} + \delta f_{n,\mathrm{reg}}\right) 
  \,\leq\, 2 |d^3\sigma| \int_p (u\cdot p) f_{\eq, n} \,,
\ee
where
\be
\label{maxvol}
  |d^3\sigma| = \sqrt{(u\cdot d^3\sigma)^2 
                - d^3\sigma \cdot d^3\sigma},
\ee
to establish an upper limit for $\Delta N_n$ in \eqref{eq:meanHadrons}: 
\be
\label{eq:meanHadronsMax}
  \Delta N_n \leq \Delta N_{n,\text{max}} 
  = 2\,|d^3\sigma|\,n_{\eq,n}.
\ee
We then sample additional particles by replacing in Eqs.~\eqref{eq:poisson}-\eqref{eq:discrete} $\Delta N_n$ with $\Delta N_{n,\text{max}}$ \cite{Pratt:2010jt,Bernhard:2018hnz}. After sampling their type and momentum (as described in the following subsection), we keep these additional particles with probability
\be
\label{eq:keep_weight}
w_\mathrm{keep}(p) = w_{d\sigma} \times w_{\delta f} = \frac{p\cdot d^3\sigma\ \Theta(p\cdot d^3\sigma)}{(u \cdot p)|d^3\sigma|} \times \frac{1}{2}\left(1 + \frac{\delta f_{n,\text{reg}}}{f_{\eq,n}}\right) \,,
\ee
where $w_{d\sigma}$ and $w_{\delta f}$ are called the flux and viscous weights whose product satisfies $0 \leq w_{\text{keep}}\leq 1$. In this way the right fraction of particles is discarded to recover, after sampling many events, the correct mean hadron number
\be
\label{eq:meanHadrons_linear}
\Delta N_\text{h}(x) = \sum_n \int_p \, p \cdot d^3\sigma \, \left(f_{\eq,n} + \delta f_{n,\text{reg}}\right) \, \Theta(p \cdot d^3\sigma).
\ee

If $f_n$ is one of the modified equilibrium distributions (\ref{jonah}) or (\ref{Mike}), no regulation \eqref{eq:df_reg} of $f_n$ is needed, i.e. the viscous weight $w_{\delta f}$ is set to 1, and the maximum hadron number is
\be
\label{eq:meanHadronsMax_mod}
\Delta N_n \leq \Delta N_{n,\text{max}} = |d^3\sigma|\,n_{R,n}
\ee
where the renormalized particle density is given by $n_{R,n} = \mathcal{Z}_\Pi\,n_{\eq,n}$ for the PTB distribution (\ref{jonah}) and by $n_{R,n} = n_n^{(1)}$ for the PTM distribution (see Eq.~(\ref{linear_density})). After sampling many events this procedure recovers the correct mean hadron number
\be
\Delta N_\text{h}(x) = \sum_n \int_p \, p \cdot d^3\sigma \, f^\mathrm{mod}_{\eq,n} \, \Theta(p \cdot d^3\sigma),
\ee
where the superscript `mod' stands generically for either PTB or PTM.

\subsection{Sampling the particle momentum}
\label{sec2c}

After sampling a hadron and its type from a freezeout cell, we sample its local-rest-frame momentum $\bp_{_\mathrm{LRF}}$ from the probability density function $Q_n(\bp) \, d^3p$, where $Q_n$ is either (suppressing the $x^\mu$ dependence)
\be
\label{eq:PDF}
  Q_n(\bp) = 
  \frac{2 |d^3\sigma| \, f_{\eq,n}(\bp)}
       {\Delta N_{n,\text{max}}}
\ee
for linearized $\delta f_n$ or 
\be
\label{eq:PDF_mod}
Q_n(\bp) = \frac{|d^3\sigma| \, f^\mathrm{mod}_{\eq,n}(\bp')}{\Delta N_{n,\text{max}}}
\ee
for the modified equilibrium distributions. We sample the momentum from Eqs.~\eqref{eq:PDF},\eqref{eq:PDF_mod} using the acceptance-rejection (AR) method. Conceptually, the method involves drawing a momentum sample from a proposal distribution $R_n(\bm{s})$,
\be
Q_n(\bp) \, d^3p = C \times R_n(\bm{s}) \, d^3s \times w_n(\bp) \,,
\ee
where $C$ is a normalization constant and $\bp = {\bm M}({\bm s})$ is some coordinate transformation, and accepting the sampled momentum with probability $w_n(\bp)$. If the sample is rejected, the procedure is repeated until a sampled momentum assignment is accepted. Once the assigned momentum for this particle has been accepted, the weight $w_\mathrm{keep}(p)$ in \eqref{eq:keep_weight} (which enforces that the momentum points outward and the viscous correction remains within the regulated range) can be calculated and used to decide whether to keep the particle or discard it. 

The proposal distribution $R_n({\bm s})$ must be chosen judiciously such that the associated weight satisfies the condition $0 \leq w_n(\bp) \leq 1$. To increase the acceptance rate the weight should also be as close to unity as possible.\footnote{%
    In our sampling routine the average acceptance rate for the momentum sampling loop is about 60\% for the modified equilibrium distributions and about half of that for the linearized $\delta f_n$ corrections because for the latter about twice as many particles than ultimately desired must be sampled to account for the factor $\frac{1}{2}$ in the viscous weight $w_{_{\delta f}}$ in Eq.~\eqref{eq:keep_weight}.}
In the following subsections, we describe the choice of $R_n({\bm s})$ and the associated weight for sampling the momenta of pions and other (heavier) hadrons from either Eq.~\eqref{eq:PDF} or \eqref{eq:PDF_mod}. 

\subsubsection{Pions with linear viscous corrections}

For pions with linear viscous corrections, it is efficient to sample the momentum from a massless Boltzmann distribution~\cite{Pratt:2014vja}
\be
\label{eq:massless}
R_n({\bm s}) \, d^3s = \exp\left(-p/T \right) \, p^2 \, dp \, d\cos\theta \, d\phi \,,
\ee
where we use the spherical coordinates ${\bm s} = (p, \cos\theta, \phi)$:
\bs
\label{eq:spherical}
\beal
& p_x = p \sin\theta \cos\phi, \\
& p_y = p \sin\theta \sin\phi, \\
& p_z = p \cos\theta \,.
\end{align}
\es
The distribution~\eqref{eq:massless} can be easily sampled using Scott Pratt's trick, which employs an additional coordinate transformation
\bs
\label{eq:PrattLight}
\beal
& p = - T \, \ln(r_1 \, r_2 \, r_3) \\
& \cos\theta = \frac{\ln(r_1/r_2)}{\ln(r_1\,r_2)} \\
& \phi = 2 \pi \left(\frac{\ln(r_1\,r_2)}{\ln(r_1\,r_2\,r_3)}\right)^2 \,,
\end{align}
\es
where the random variables $(r_1,r_2,r_3) \in (0,1]$. The massless Boltzmann distribution~\eqref{eq:massless} can then be rewritten as (ignoring constant factors)
\be
R_n({\bm s}) \, d^3s = dr_1 dr_2 dr_3 \,.
\ee
As a result, one can simply sample $(r_1,r_2,r_3)$ uniformly and apply the transformations~\eqref{eq:PrattLight} and~\eqref{eq:spherical} for the momentum components $\bm p$.

The weight factor for pions is 
\be
\label{eq:pion_weight}
w_n(\bm p) = \frac{\exp\left(p/T\right)}{\exp\left(E/T\right) - 1}
\ee
where $E = \sqrt{p^2 + m_n^2}$. One finds that the thermal weight $0 \leq w_n(p) \leq 1$ for all values of $p$ when $m / T > 0.8554$. For the lightest pion mass, this corresponds to $T < T_\text{max} = 157.8$ MeV. This is acceptable because particlization typically occurs several MeV below the pseudocritical temperature $T_c = 155$ MeV.\footnote{For situations where the switching temperature $T_\text{sw} \geq T_\text{max}$, one can renormalize the thermal weight by its maximum value, which we solve for numerically.} 

\subsubsection{Heavy hadrons with linear viscous corrections}

For heavy hadrons, we sample the momentum from the Boltzmann distribution~\cite{Pang:2018zzo,Pratt:2014vja}
\be
\label{eq:classical}
R_n({\bm s}) \, d^3s = \exp\left(b_n \alpha_B{-}E/T\right) \, p^2 \, dp \, d\cos\theta \, d\phi \,.
\ee
We use an intermediate variable transformation $k = E{-}m_n$, with $k$ being the kinetic energy, to rewrite Eq.~\eqref{eq:classical} as (omitting constant factors)
\be
R_n({\bm s}) \, d^3s = \,\frac{p}{E} \, S_n({\bm k}) \, d^3k, 
\ee
where ${\bm k} \equiv (k,\cos\theta,\phi)$ and
\be
\label{eq:sampleK}
S_n({\bm k}) \, d^3k = \, \exp\left(-k/T \right) \, \left(k^2 + 2km_n + m_n^2\right) \, dk \, d\cos\theta \, d\phi \,.
\ee
Thus, we can sample the kinetic energy and angles from the distribution $S_n({\bm k})$ (moving the factor $p/E$ over to the weight $w_n({\bm p})$) to compute the energy $E$, radial momentum $p = \sqrt{E^2{-}m_n^2}$, and momentum components $p_i$ with Eq.~\eqref{eq:spherical}. We write Eq.~(\ref{eq:sampleK}) as the sum of three distributions, 
\begin{equation}
\label{Si}    
  S_n({\bm k}) \, d^3k = 
  \Bigl(S_1({\bm k}) + 2 m_n S_2({\bm k}) 
        + m_n^2 S_3({\bm k})\Bigr)\, d^3k,
\end{equation}
with
\bs
\label{eq:heavy3}
\beal
& S_1({\bm k}) \, d^3k = \exp\left(-k/T \right) \, k^2 \, dk \, d\cos\theta \, d\phi, \\
& S_2({\bm k}) \, d^3k = \exp\left(-k/T \right) \, k \, dk \, d\cos\theta \, d\phi, \\
& S_3({\bm k}) \, d^3k = \exp\left(-k/T \right) \, dk \, d\cos\theta \, d\phi \,.
\end{align}
\es
The first distribution (\ref{eq:heavy3}a) is identical to the massless Boltzmann distribution~\eqref{eq:massless} after replacing the variable $p$ with $k$, so we can sample it with the same technique. For the second distribution (\ref{eq:heavy3}b), we use the coordinates
\beal
\label{eq:PrattSemi}
 k = - T \, \ln(r_1 \, r_2), \quad
 \phi = \frac{2 \pi \, \ln(r_1)}{\ln(r_1\,r_2)} \,,
\end{align}
where $(r_1, r_2) \in (0,1]$. One can check that $S_{2}({\bm k}) \, d^3k \sim dr_1 \, dr_2 \, d\cos\theta$, which means we can sample $(r_1, r_2, \cos\theta)$ uniformly. For the third distribution (\ref{eq:heavy3}c), we substitute 
\be
k = - T\,\ln(r_1) \,,
\ee
where $r_1 \in (0,1]$ and sample $(r_1, \cos\theta, \phi)$ uniformly since $S_{3}({\bm k}) \, d^3k \sim$ $dr_1$ $d\cos\theta$ $d\phi$.

Each time we draw a momentum sample, instead of drawing it from the total distribution \eqref{eq:sampleK} we draw it from one of these three simpler distributions, selected randomly with probabilities $(I_1,I_2,I_3)/I_\text{tot}$ where
\bs
\label{eq:integratedWeights}
\beal
& I_1 = \int_0^\infty dk \, k^2 \, \exp\left(-k/T \right) = 2 T^3 \\
& I_2 = 2 m_n \int_0^\infty dk \, k \, \exp\left(-k/T \right) = 2 m_n T^2 \\
& I_3 = m_n^2 \int_0^\infty dk \exp\left(-k/T \right) = m^2_n T
\end{align}
\es
and $I_\text{tot} = I_1 + I_2 + I_3$ \cite{Pang:2018zzo, Pratt:2014vja}. After many draws this ensures that, on average, the momenta have been drawn from the desired distribution \eqref{eq:sampleK}.

The weight for heavy hadrons is then chosen as
\be
w_n(\bm p) = \frac{p}{E} \, \frac{\exp\left(E/T - b_n \alpha_B\right)}{\exp\left(E/T - b_n \alpha_B\right) + \Theta_n}.
\ee
For baryons ($\Theta_n{\,=\,}1$) this thermal weight satisfies $0 \leq w_n(p) \leq 1$ for all values of $p$. For mesons ($\Theta_n{\,=\,}{-}1$) the weight condition holds for $m/T > 1.008$.\footnote{%
    For pions and heavy mesons, the thermal weight $w_n(p)$ can exceed unity for situations where the electric and net-strangeness chemical potentials are nonzero. For this reason, we leave the generalization to sampling particles with nonzero $(\mu_Q, \mu_S)$ to future work.}
Since even the mass of the lightest of them, the kaon, is three times larger than the typical switching temperature, the weight condition $0 \leq w_n(p) \leq 1$ is satisfied for all heavy hadrons. 

\subsubsection{Modified equilibrium distribution}

To sample momenta from the PTB or PTM modified equilibrium distributions, we use the momentum transformation \eqref{rescaling} or \eqref{eq:transform_Mike} to rewrite Eq.~\eqref{eq:PDF_mod} as (ignoring constant factors)
\be
\label{eq:modified_PDF_2}
  Q_n(\bp) \, d^3p = 
  \left( 1-\frac{q_i A^{-1}_{ij} p'_j}
                {\sqrt{{p'}^2 + m_n^2}}
  \right)
  f_{\eq,n}^{\text{mod}}(\bm{p'}) \, d^3p^\prime \,,
\ee
where $q_i$ is given by~\eqref{eq:baryon_Mike} ($q_i = 0$ for the PTB distribution). One can then sample the modified momentum components $\bm{p'}$ from $f_{\eq,n}^{\text{mod}}(\bm{p'})$ and apply the viscous transformation \eqref{rescaling} or \eqref{eq:transform_Mike} for $\bm{p}$ \cite{Pratt:2010jt,Bernhard:2018hnz}. For pions, we sample $\bm{p'}$ from the modified massless Boltzmann distribution
\be
\label{eq:massless_modified}
R_n(\bm s) \, d^3s = \exp\left(-p'/T' \right) \, (p^\prime)^2 \, dp' \, d\cos\theta' \, d\phi' \,,
\ee
where we use the \textit{modified} spherical coordinates ${\bm s} = (p^\prime, \cos\theta^\prime, \phi^\prime)$:
\bs
\label{eq:spherical_prime}
\beal
& p'_x = p' \sin\theta' \cos\phi' \,, \\
& p'_y = p' \sin\theta' \sin\phi' \,, \\
& p'_z = p' \cos\theta' \,.
\end{align}
\es
It is straightforward to sample ($p^\prime$, $\cos\theta^\prime$, $\phi^\prime$) with Scott Pratt's trick. The modified weight for pions is
\be
   w_n(\bm p) = w_{q} \times
   \frac{\exp(p'/T')}
        {\exp(E'/T') - 1}
\ee
where
\be
\label{eq:diffusion_weight}
  w_{q} = \frac{1 - q_i A^{-1}_{ij} p'_j/E'}
               {1 + \sqrt{A^{-1}_{rs} A^{-1}_{rs}\, {\bm q}^2}}
\ee
is the baryon diffusion weight and $E' = \sqrt{{p'}^2 + m_n^2}$.\footnote{%
    Here we assume that the numerator of the diffusion weight \eqref{eq:diffusion_weight} is positive since we require a positive Jacobian determinant \eqref{eq:Jacobian}. Future work will address whether the PTM distribution may break down for large baryon diffusion currents.}
It is important to note that the modified temperature $T'$ in the PTM distribution increases with negative bulk pressure, which could make the pion effectively too light (i.e. lead to $m/T' < 0.8554$). If the bulk pressure is too large, the thermal weight can be renormalized by its maximum value (as long as it is finite) to ensure that it stays below unity.

For heavy hadrons, we sample $\bm{p'}$ from the modified Boltzmann distribution
\be
   R_n({\bm s}) \, d^3s = \exp
   \left( b_n \alpha'_B - E'/T'\right) \, 
   {p'}^2 dp' \, d\cos\theta'\, d\phi' \,,
\ee
where (in the PTM distribution) $\alpha'_B$ is the modified baryon chemical potential. After substituting $k' = E'{-}m_n$, we have
\be
\begin{split}
   R_n({\bm s}) \, d^3s 
   = & \, \frac{p'}{E'} \, S_n({\bm k'}) \, d^3k' \\
   = & \, \frac{p'}{E'} \, \exp(-k'/T') \, 
     \bigl({k'}^2 + 2k' m_n + m_n^2\bigr) \, dk^\prime \, d\cos\theta^\prime \, d\phi^\prime \,.
\end{split}
\ee
The procedure for sampling ${\bm k'} = (k',\, \cos\theta',\, \phi')$ is analogous to sampling $\bm{k}$ from Eq.~\eqref{eq:sampleK}. The modified weight for heavy hadrons is then
\be
  w_n(\bm p) = w_q \times \frac{p'}{E'} \, 
  \frac{\exp(E'/T' - b_n \alpha'_B) }
       {\exp(E'/T' - b_n \alpha'_B) + \Theta_n} \,.
\ee
%

\begin{figure}[t!]
\begin{center}
  \includegraphics[width=0.85\linewidth]{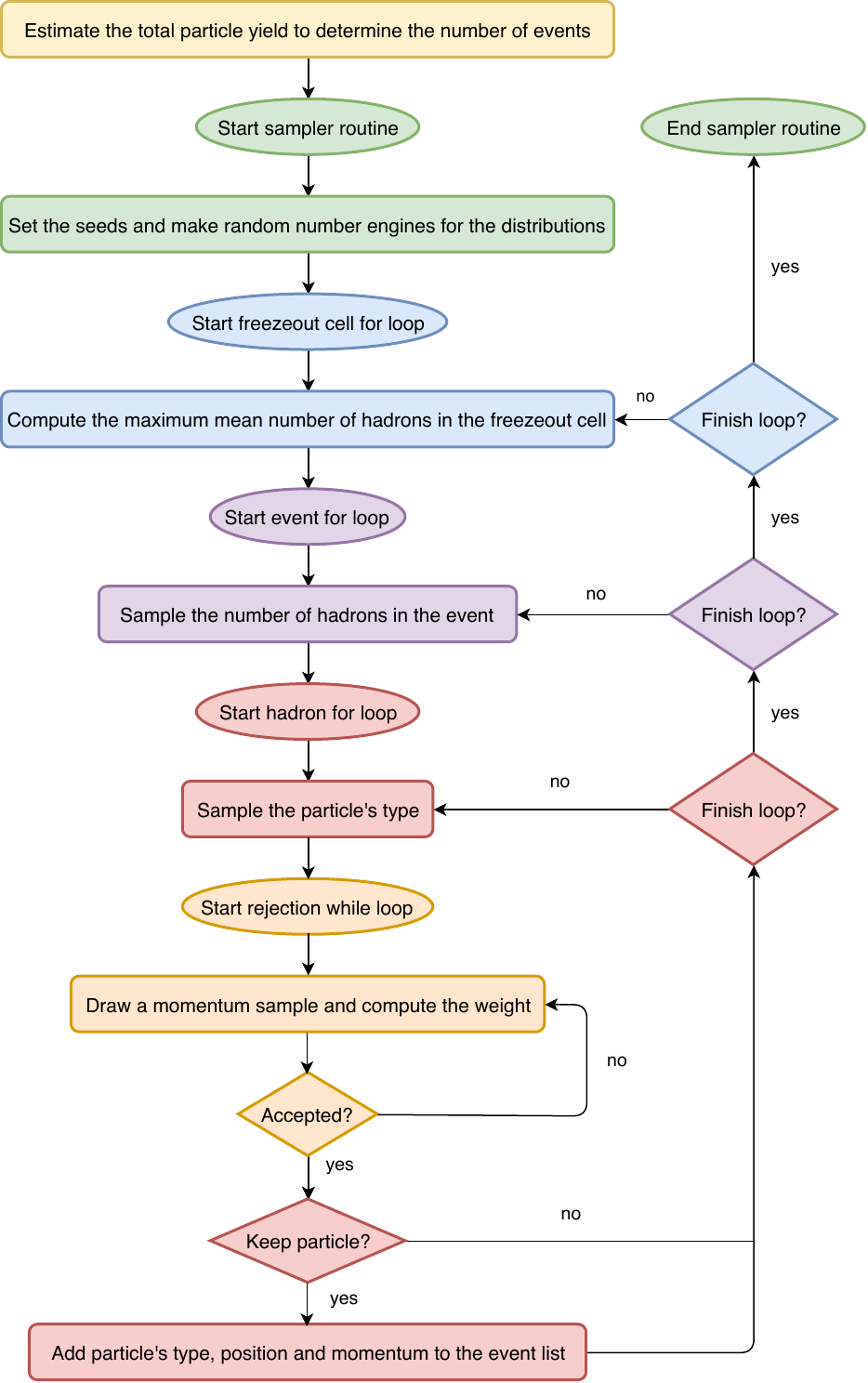}
\end{center}
  \caption{(Color online)
    The program flow chart of the particle sampler routine in \I.
    \label{F1}
    }
\end{figure}

\subsection{Program flow chart}
\label{sec5.3}

Figure~\ref{F1} shows the program flow chart of the particle sampling routine in iS3D. Here we summarize the steps of the sampling procedure: 
\begin{enumerate}
    \item Before calling the sampler routine, we estimate the total particle yield per collision event (without including the effects of outflow or regulating $\delta f_n$)
    \be
    N_\text{yield} \approx \sum_n \int_\Sigma \int_p  (u \cdot d^3\sigma) (u\cdot p) \, \left(f_{\eq,n} + \delta f_n\right)
    \ee
    to determine the number of sampled events $N_\text{event} = N_\text{sampled} / N_\text{yield}$ needed to accumulate the desired statistics of approximately $N_\text{sampled}$ particles from the switching hypersurface ($N_\text{sampled}$ is a user input parameter).

    \item We call the sampler routine, initializing the seeds and random number engines for each of the three distributions $P(N)$ (\ref{eq:poisson}), $D_n$ (\ref{eq:discrete}) and $Q_n({\bm p})$ (see Sec.~\ref{sec2c}).
    \item We loop over the freezeout cells. For each freezeout cell, we evaluate the position $x^\mu$, surface volume element $d^3 \sigma_\mu$, hydrodynamic quantities ($u^\mu$, $T$, $\mathcal{E}$, $\mathcal{P}_\text{eq}$, $\pi^{\mu\nu}$, $\Pi$), and $\delta f_n$ coefficients; we skip over freezeout cells with negative timelike volumes (i.e. $u \cdot d^3\sigma < 0$). Then, we construct the hadron number and hadron species distributions $P(N)$ and $D_n$ by computing the maximum number of each hadron species emitted from the freezeout cell \eqref{eq:meanHadronsMax} or \eqref{eq:meanHadronsMax_mod}.

    \item For a given freezeout cell, we loop over the events. For each event, we sample the number of hadrons from the distribution $P(N)$. It is efficient to nest the event for-loop in this way because we only need to access the freezeout cell information and compute the max hadron number once. 

    \item For a given event, we loop over the sampled hadrons. For each hadron, we sample its species from the distribution $D_n$. Next, we sample the particle's local-rest-frame momentum $\bp_{_\mathrm{LRF}}=(p_x,p_y,p_z)$ from the distribution $Q_n(\bp)$ using the AR method. We keep the particle with probability $w_\text{keep}$ and, if accepted,  we compute its lab frame momentum from (see footnote \ref{fn10})
    \be
    \label{lab_mom}
      p^\mu = E u^\mu + p_x X^\mu + p_y Y^\mu + p_z Z^\mu 
    \ee
    and also set the particle's lab frame position to that of the freezeout cell. Finally, we append the particle to the list of particles sampled in the event.

    \item After the sampler routine is finished, we write the particle data list of each event to file. This file follows the OSCAR format \cite{OSCAR} to allow for integration with hadronic afterburners such as URQMD and SMASH. 
\end{enumerate}

For the special case of a (2+1)-dimensional hydrodynamic switching surface with longitudinal boost invariance (as used in the performance tests presented in this document) the sampler routine is modified in two ways \cite{Bernhard:2018hnz}: First, we note that for boost-invariant switching surfaces, the surface finder CORNELIUS \cite{Huovinen:2012is} assumes by default a longitudinal extension of one unit of space-time rapidity, $\Delta\eta_s=1$. When computing the mean number of hadrons according to Eq.~(\ref{eq:meanHadrons}) we multiply the midrapidity surface volume element $d^2\sigma_i$ by a factor $2\,y_\mathrm{max}$ to obtain $\Delta\eta_s=2\,y_\mathrm{max}$ (the rapidity cutoff $y_\mathrm{max}$ is a parameter set by the user). Second, after sampling the LRF momentum and accepting the particle, we compute its lab frame momentum from Eq.~(\ref{lab_mom}). Expressing the Milne lab frame momentum components as $p^\tau=m_T\cosh(y_p{-}\eta_s)$ and $p^\eta = (m_T/\tau) \sinh(y_p{-}\eta_s)$, the particle's momentum rapidity $y_p$ relative to its space-time rapidity $\eta_s$ is $y_p{-}\eta_s =  \tanh^{-1}\left(\tau p^\eta / p^\tau\right)$. We then generate a boost-invariant (i.e. constant) distribution $\frac{dN_n}{dy_p}$ over the range $y_p\in[-y_\mathrm{max}, y_\mathrm{max}]$  by sampling $y_p$ uniformly within the interval $[-y_\text{max}, y_\text{max}]$. This also yields a space-time rapidity distribution $\frac{dN_n}{d\eta_s}$ after assigning the particle the space-time rapidity
\be
\label{kinematic}
  \eta_s =  y_p - \tanh^{-1}\left(\frac{\tau p^\eta}{p^\tau}\right)
\ee
The resulting pair $(y_p,\,\eta_s)$ is attached to the accepted particle before it is written to the particle data list for the event. For sufficiently large $y_\text{max}$, the uniform sampling of $y_p$, followed by the kinematic constraint (\ref{kinematic}), ensures after event-averaging a constant (i.e. perfectly boost-invariant) and correctly normalized mean yield $\frac{dN_n}{dy_p}$ in the range $y_p\in[-y_\text{max}, y_\text{max}]$, combined with a space-time rapidity distribution $\frac{dN_n}{d\eta_s}$ that is approximately constant except for edge effects localized near $\eta_s{\,=\,}\pm y_\mathrm{max}$. 

\section{Particle sampler performance}
\label{S6}

In this section we test the performance of our particle sampler by comparing the event-averaged particle space-time distributions and momentum spectra to the positive-definite Cooper Frye formula \eqref{eq:sampledCFF} for the central and non-central Pb-Pb collision systems described in Sec.~\ref{S3}. Our hadron resonance gas consists of the $N_R = 444$ hadron species that can be propagated in SMASH \cite{Weil:2016zrk}. For each hypersurface from the (2+1)-d hydrodynamic model we multiply the volume by a factor 10 (by setting $y_\text{max}{\,=\,}5$) and sample a total of approximately $N_\text{sampled} \approx 10^{11}$ particles. This corresponds to sampling between five to ten million events, depending the kind of hypersurface and choice for $\delta f_n$. When testing the sampler, we bin the particles in position and momentum grids during runtime to construct the sampled distributions, rather than accumulating all of the sampled particle data for later processing in a separate analysis. This avoids running into RAM limitations and file I/O bottlenecks that result from generating so many particles.

\subsection{Central Pb-Pb collision}
\label{S6.1}

\begin{figure}[t]
\vspace*{-20mm}
\makebox[\textwidth][c]{\includegraphics[width=\textwidth]{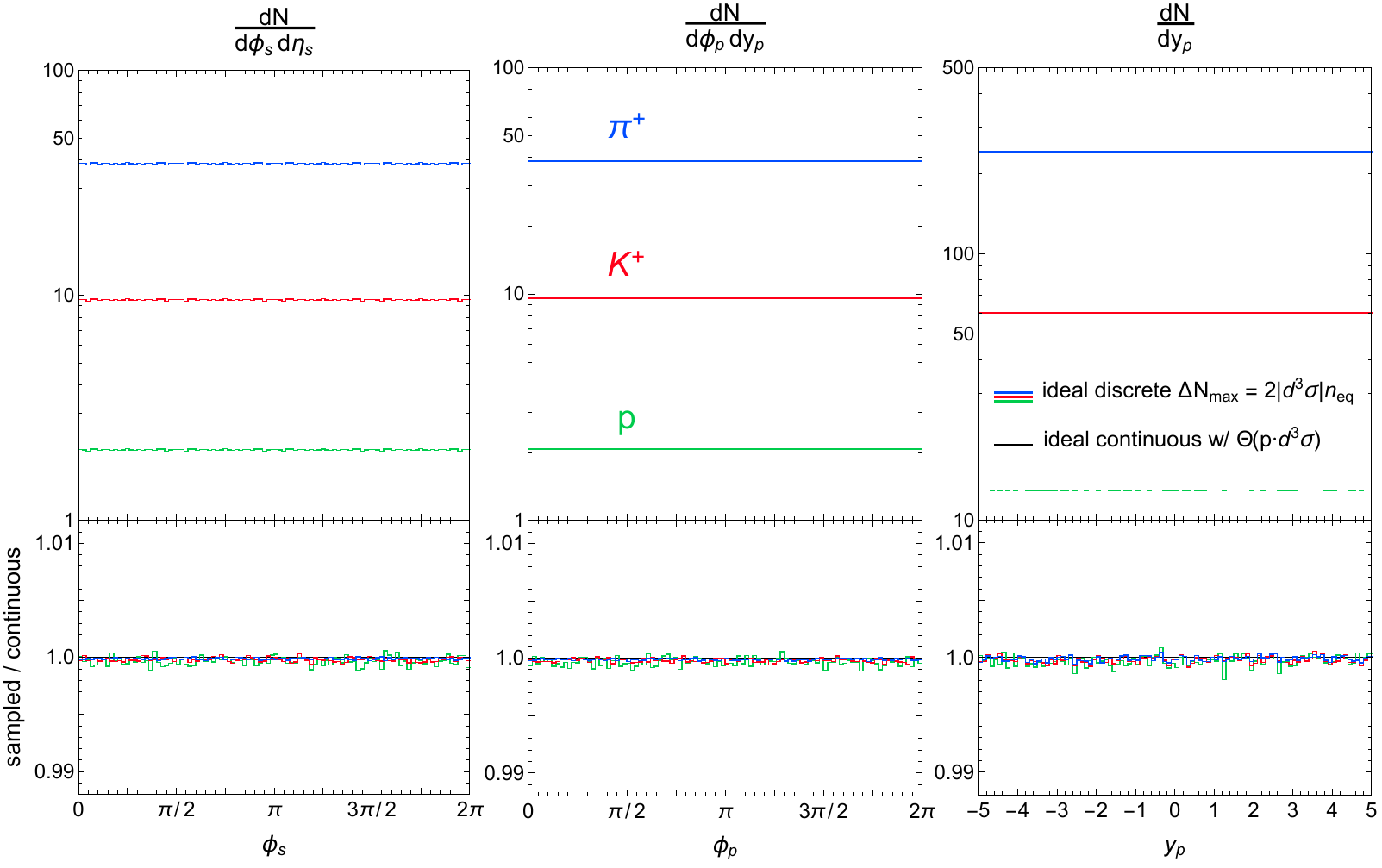}}
\caption{(Color online)
The azimuthal and rapidity distributions of $(\pi^+,K^+,p)$ from the 2+1d central Pb-Pb collision with a $\zeta/\mathcal{S}$ peak temperature of $T_p = 180$ MeV. The $\delta f_n$ correction was set to zero (\textit{ideal}). The bin widths used are $\Delta \phi_s = \Delta \phi_p = 0.02\, \pi$ and $\Delta y_p = 0.1$. Both the sampled (solid colored) and continuous (solid black) distributions include the outflow correction $\Theta(p\cdot d^3\sigma)$ to the Cooper-Frye formula. The bottom panels show the ratios between the sampled and continuous distributions. 
\label{Fideal_dN_dy}
}
\end{figure}

In the first test, we sample the Cooper-Frye formula for the central Pb-Pb collision with a $(\zeta / \mathcal{S})(T)$ that peaks at a temperature $T_p = 180$\,MeV. We construct the discrete transverse momentum spectra~\eqref{eq:central_pT_spectra} by binning the sampled particles in a uniform $p_T$-grid with width $\Delta p_T = 0.03$ GeV, averaging over the events and rapidity. For the sampled temporal and radial distributions (\ref{eq:time_formula}, \ref{eq:radial_formula}), we bin the particles in the same $(\tau, r)$-grid as the one used in Sec.~\ref{S4}. We have also verified that the event-averaged particle distributions are azimuthally symmetric in position and momentum space and are longitudinally boost-invariant, as shown in Figure~\ref{Fideal_dN_dy}.  

\begin{figure}
\makebox[\textwidth][c]{\includegraphics[width=\textwidth]{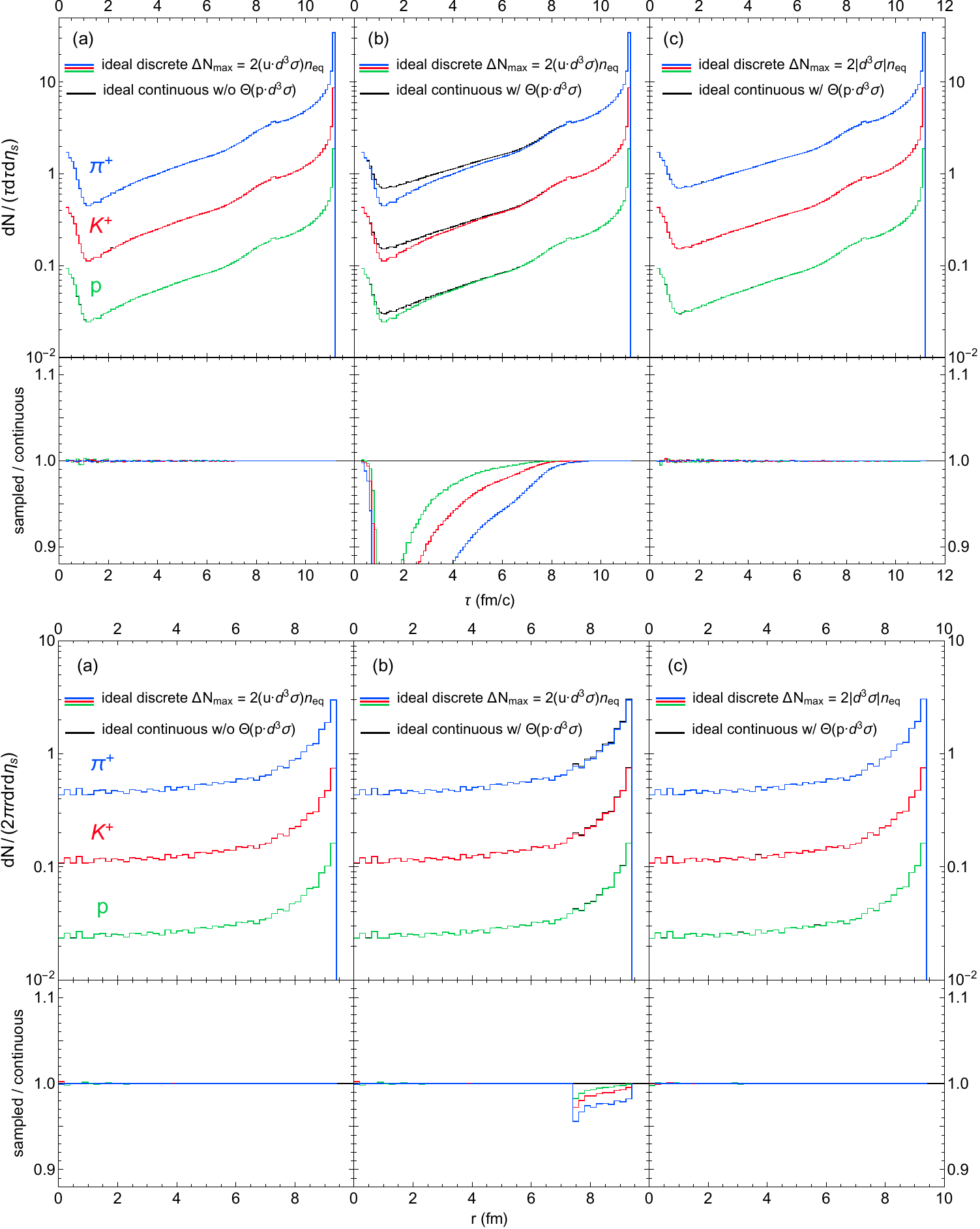}}
\caption{(Color online)
The temporal (top panels) and radial distributions (bottom panels) of $(\pi^+,K^+,p)$ for the (2+1)-d central Pb-Pb collision with a $\zeta/\mathcal{S}$ peak temperature of $T_p = 180$\,MeV. The $\delta f_n$ correction was set to zero (\textit{ideal}). The sampled distributions (solid colored) are generated without (a,b) or with (c) the outflow correction to the mean hadron number. The continuous distributions (solid black) are computed without (a) or with (b,c) the $\Theta(p \cdot d^3\sigma)$ function. The bottom subpanels show the ratios between the sampled and continuous distributions for each of the comparisons. 
\label{Fidealspace-time}
}
\end{figure}

\begin{figure*}[t]
\makebox[\textwidth][c]{\includegraphics[width=\textwidth]{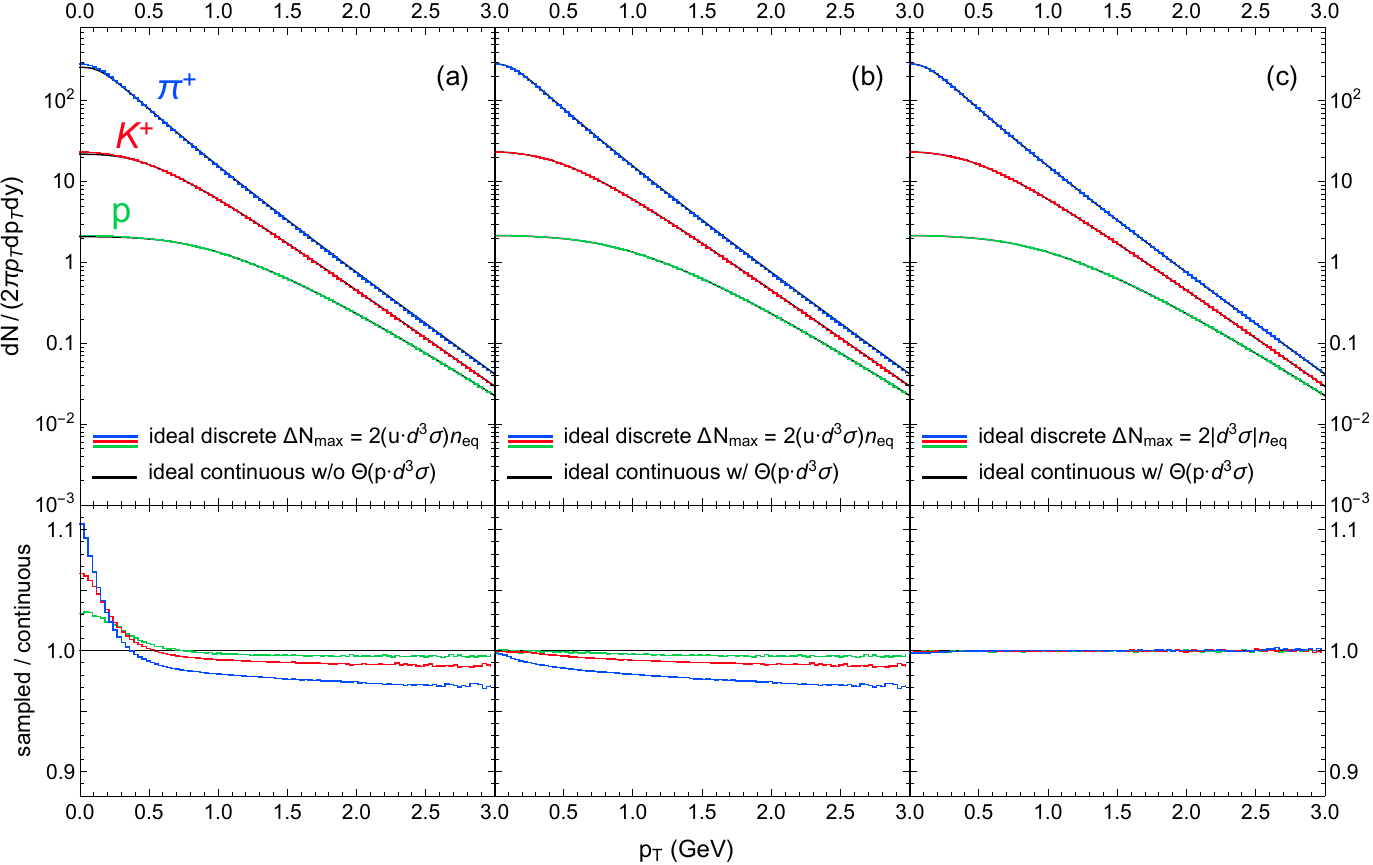}}
\caption{(Color online)
The same comparisons as in Fig.~\ref{Fidealspace-time} but for the azimuthally-averaged transverse momentum spectra \eqref{eq:central_pT_spectra}.
\label{Fidealspectra}
}
\end{figure*}

\subsubsection{The effect of particle outflow}
\label{S6.1.1}

In this subsection we study the effects of the outflow correction, implemented by the function $\Theta(p \cdot d^3\sigma)$, on the space-time distributions and momentum spectra. For simplicity, and without loss of insight, we set $\delta f_n = 0$ in this comparison.\footnote{%
    We label space-time distributions and momentum spectra without any $\delta f_n$ correction as \textit{ideal}, but this does not imply $\eta/\mathcal{S}$ and $\zeta/\mathcal{S}$ are set to zero during the viscous hydrodynamic simulation. We only turn off the $\delta f_n$ correction on the switching surface.}
Fig.~\ref{Fidealspace-time} shows the temporal and radial distributions of ($\pi^+, K^+, p$). In Fig.~\ref{Fidealspace-time}a,b we sample the particles without the outflow correction to the mean hadron number (i.e. $\Delta N_n = (u \cdot d^3\sigma)\, n_{\text{eq},n}$).\footnote{%
    If the outflow effect on the particle yield is not included, we replace the freezeout cell volume $|d^3\sigma|$ with $(u \cdot d^3\sigma)$ in Eqs.~\eqref{eq:meanHadronsMax} and \eqref{eq:meanHadronsMax_mod}. In additon, we move the flux weight $w_{d\sigma}$ from $w_\text{keep}(p)$ to the thermal weight $w_n(p)$.}
We then compare the resulting sampled space-time distributions to the continuous ones computed from the Cooper-Frye formula with and without the $\Theta(p \cdot d^3\sigma)$ function. Clearly, the continuous method without the $\Theta(p \cdot d^3\sigma)$ function also yields $\Delta N_n = (u \cdot d^3\sigma)\, n_{\text{eq},n}$ for each freezeout cell. Thus, the sampled and continuous distributions are in very good agreement for the first case (Fig.~\ref{Fidealspace-time}a). In the second case (Fig.~\ref{Fidealspace-time}b), where this time we compute the Cooper-Frye formula with the $\Theta(p \cdot d^3\sigma)$ function, the continuous distribution is greater than the sampled one at early times and at large radii. This is because in these space-time regions the hypersurface elements are spacelike. The sampled and continuous distributions are only in agreement for the upper part of the hypersurface in Fig.~\ref{FHydro} ($\tau \gtrsim 9\,\text{fm}/c,\, r \lesssim 7\,\text{fm})$, where the hypersurface elements are timelike and $\Theta(p \cdot d^3\sigma) = 1$ has no effect. For the third case (Fig. \ref{Fidealspace-time}c) we sample particles from freezeout cells of maximum volume $|d^3\sigma|$ defined in Eq.~(\ref{maxvol}), which reproduces the outflow correction to the particle yield after keeping particles with probability $w_\text{keep}$. Correspondingly, the resulting sampled distribution is in excellent agreement with the continuous distribution computed with the $\Theta(p \cdot d^3\sigma)$ function.

In Fig.~\ref{Fidealspectra} we make the same comparisons for the transverse momentum spectra. In Fig.~\ref{Fidealspectra}a both the sampled and continuous spectra ignore the outflow effect on the mean number of hadrons emitted from the hypersurface; thus, they have the same particle yields. However, sampling the momentum must still be done with the $\Theta(p \cdot d^3\sigma)$ function; this gives preference to the emission of softer particles over harder ones from the spacelike regions of the hypersurface. As a result, the sampled spectra are softer than the continuous spectra computed without the $\Theta(p \cdot d^3\sigma)$ function; the discrepancy in the low $p_T$ region is as high as 10\% for pions. This implies that in this mode the particle sampler is not able to conserve energy and momentum, even if the sampled particle yield matches the one given by the original Cooper-Frye Formula. For the second case (Fig.~\ref{Fidealspectra}b) we compare the sampled spectra without the outflow effect on the particle yield to the continuous spectra computed with the $\Theta(p \cdot d^3\sigma)$ function. The sampled and continuous spectra now show better agreement in shape, but the sampled spectra still underestimate the continuous particle yield by a few percent. In the third case (Fig.~\ref{Fidealspectra}c) both the sampled and continuous spectra include the outflow correction to the particle yield. Now that the sampled and continuous methods are consistent with each other, we obtain nearly perfect agreement between the spectra. We conclude that in order to conduct high precision tests on the particle sampler one must include the outflow effect on the mean hadron number in Eq.~\eqref{eq:meanHadrons}.

\subsubsection{Small regulated viscous corrections}
\label{central_180}

\begin{figure}[htbp]
\vspace*{-20mm}
\makebox[\textwidth][c]{\includegraphics[width=\textwidth]{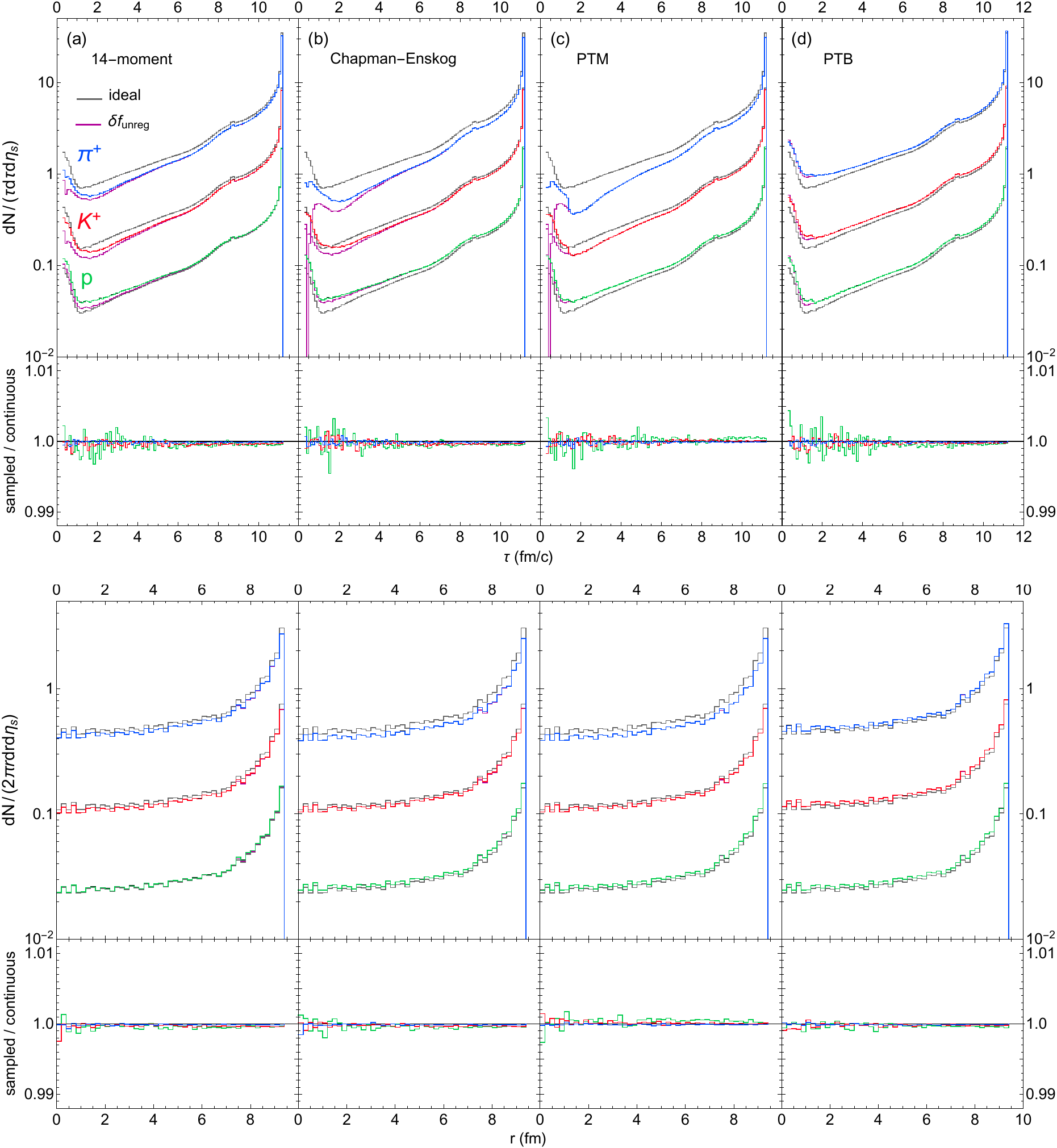}}
\caption{(Color online)
The sampled (solid blue ($\pi^+$), red ($K^+$) and green ($p$)) and continuous (solid black) temporal and radial distributions with both regulated shear and bulk $\delta f_n$ corrections to the \textit{ideal} (solid gray) distributions for a (2+1)-d central Pb-Pb collision with a $\zeta/\mathcal{S}$ peak temperature of $T_p = 180$ MeV. The space-time distributions with an unregulated $\delta f_n$ correction (solid purple) are also shown. In each panel the lower subpanel shows the ratio between the sampled and continuous distributions with regulated viscous corrections. 
\label{F_df_space-time_180}
}
\end{figure}

We now add both the regulated shear and bulk viscous corrections, along with the outflow correction, to the sampled space-time and momentum distributions. Fig.~\ref{F_df_space-time_180} shows the temporal and radial distributions of $(\pi^+, K^+, p)$ (color coded as seen in the figure) computed with each of the four $\delta f_n$ corrections described in Sec.~\ref{S2}. Compared to the \textit{ideal} space-time distributions with the outflow correction ($\delta f_n = 0$, gray lines), the particle production computed with the 14-moment approximation, Chapman-Enskog expansion and PTM distribution decreases for pions and kaons while it increases for protons; this is mainly due to the form of their bulk viscous corrections. For the PTB distribution, the renormalization factor $z_\Pi$ grows with negative bulk pressure, increasing the particle production of all species by the same factor.

We also compare the regulated space-time distributions to the continuous ones computed with an unregulated $\delta f_n$ correction (purple lines) (both include the particle outflow effect). One sees that the regulated temporal distributions of the 14-moment approximation and Chapman-Enskog expansion are greater than the unregulated ones at early times as a result of enforcing positivity. This is primarily due to the large hydrodynamic gradients of the fireball at early times, leading to the regulation of large bulk viscous corrections.\footnote{%
    The shear viscous correction $\delta f_{\pi,n}$ is also prone to regulation if the isotropic part of distribution function $f_{\eq,n} + \delta f_{\Pi,n}$ is close to the regulation bounds.} 
At later times, the bulk viscous pressure quickly dies down because in this example it peaks far away from the switching temperature $T_\text{sw} = 150$\,MeV (see Fig.~\ref{FHydro}a). As a result, the regulated temporal distributions start to converge to the unregulated ones. The regulated radial distributions are also slightly higher at around $r \sim 7.5 - 8$ fm; this is correlated to the regulation at early times. The modified equilibrium distributions are also regulated at early times ($\tau \lesssim$\,1.5 fm/$c$) since the viscous corrections are so large that we need to switch to a linearized $\delta f_n$ correction, which is regulated. Once we are able to transition to a modified equilibrium distribution, the regulation effects vanish. Compared to the linearized $\delta f_n$ corrections, the regulation has very little effect on the PTM and PTB space-time distributions.

\begin{figure*}[t]
\makebox[\textwidth][c]{\includegraphics[width=\textwidth]{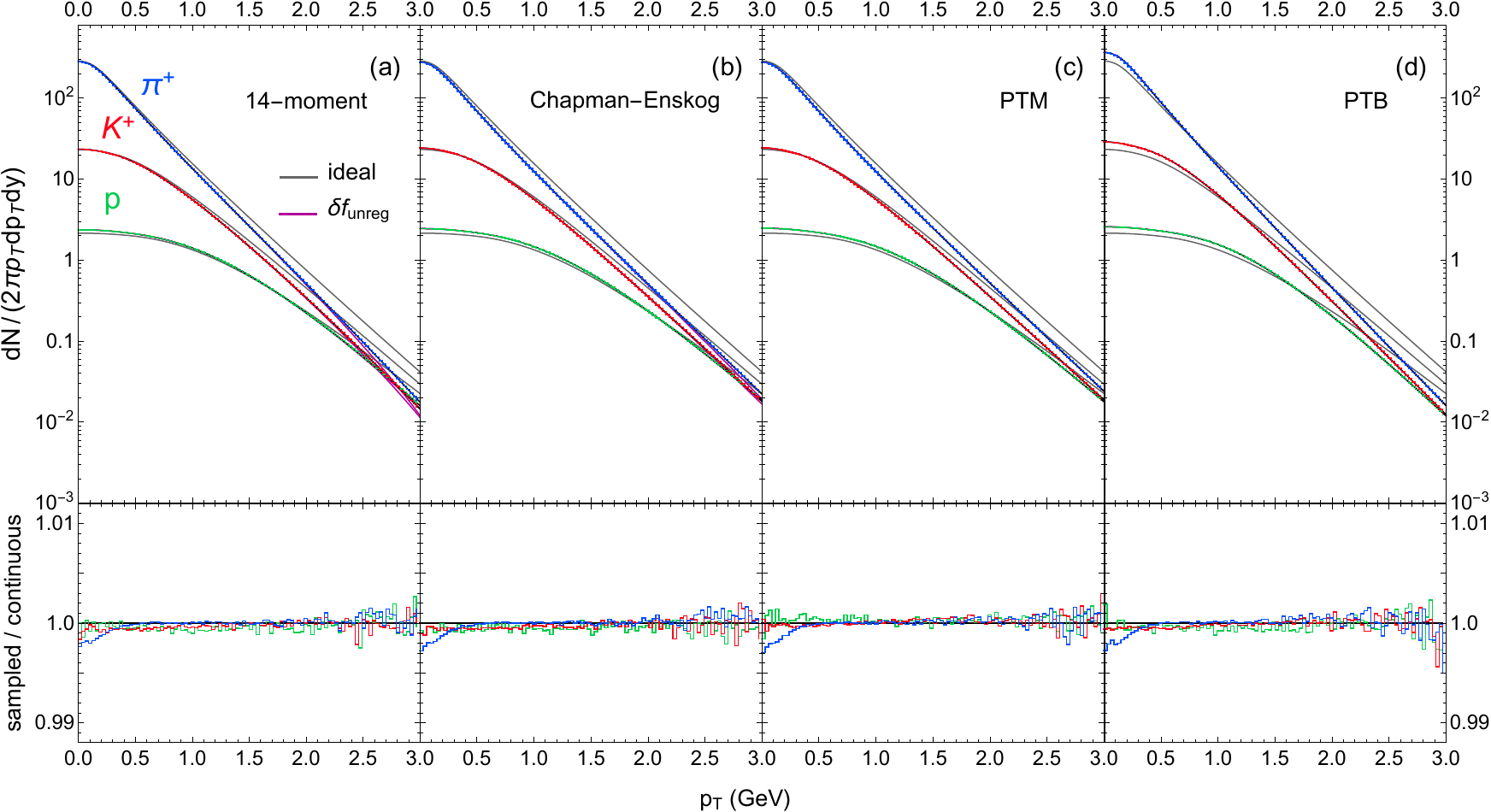}}
\caption{(Color online)
The same comparisons as in Fig.~\ref{F_df_space-time_180} but for the azimuthally-averaged transverse momentum spectra.
\label{F_df_spectra_180}
}
\end{figure*}

Finally, we compare the sampled space-time distributions to the regulated distributions. One sees that there is excellent convergence to the continuous distributions (less than 0.5\% error for all values of $(\tau, r)$). One also notices that the sampled-to-continuous ratios fluctuate slightly above (or below) unity at late times and large radii. The reason for this slight discrepancy is unclear; nevertheless a 0.05 - 0.1\% error is more than satisfactory.

Figure~\ref{F_df_spectra_180} shows the corresponding transverse momentum spectra for each $\delta f_n$ method. In general, the bulk viscous pressure softens the slope of the ideal spectra while the shear stress stress counteracts this by flattening it. Here, the bulk viscous correction exceeds the shear correction, resulting in an overall softening of the $p_T$ spectra. One notices that the sampled spectra of the 14-moment approximation and Chapman-Enskog expansion are regulated at large momentum ($p_T \gtrsim 2$\,GeV for pions and kaons); this is because the 
regulation limits the strength of the bulk viscous correction at high $p_T$. The regulation has virtually no effect on the modified spectra since the changes to the particle production in Fig.~(\ref{F_df_space-time_180}c-d) were negligible. Once again, we find nearly perfect agreement between the sampled particle spectra and the regulated continuous spectra. The sampled pion spectra, however, slightly dips below the continuous spectra at low values of $p_T$. This effect is caused by finite bin widths. Since the pion spectra is strongly concave at low $p_T$, the average spectra over each of these $p_T$ bins is lower than the midpoint value from Jensen's inequality. One can eliminate this effect simply by using narrower $p_T$ bins in this region.

\subsubsection{Large regulated viscous corrections}
\label{S6.1.3}

In this subsection we sample the hypersurface from the central Pb-Pb collision with a $(\zeta / \mathcal{S})(T)$ that peaks at a temperature of $T_p$ = 155 MeV (see Fig.~(\ref{FHydro}b)). In this scenario, the bulk viscous pressure peaks close to the switching temperature $T_\text{sw} = 150$ MeV, resulting in a much larger bulk viscous correction than in the preceding subsection. The nonlinear shear-bulk coupling terms in the $2^{\text{nd}}$ order viscous hydrodynamic equations then also increases the magnitude of the shear stress correction. 
%
\begin{figure}[!t]
\vspace*{-20mm}
\makebox[\textwidth][c]{\includegraphics[width=\textwidth]{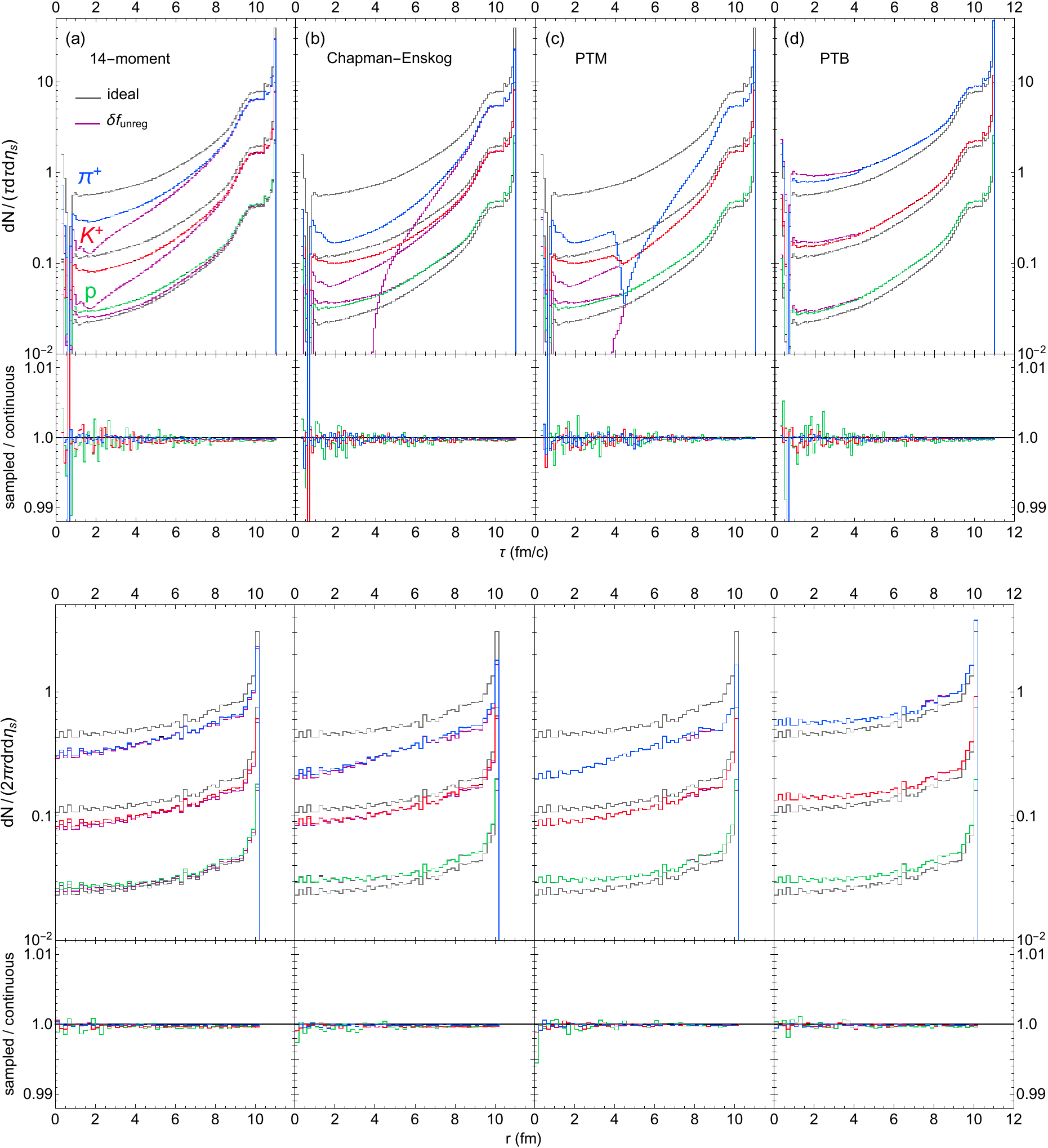}}
\vspace*{-5mm}
\caption{(Color online)
The same as Fig.~\ref{F_df_space-time_180} but for the central Pb-Pb collision with a peak temperature of $T_p = 155$ MeV for $\zeta/\mathcal{S}$.
\label{F_df_space_155}
}
\end{figure}
%
Figure~\ref{F_df_space_155} shows the resulting space-time distributions. The effects of the linearized $\delta f_n$ regulation on the particle production rates are much more significant and they have a longer duration, especially for the pion and kaon production rates. For the unregulated Chapman-Enskog $\delta f_n$ correction, the pion production rate is even negative until about $\tau \approx 4$ fm/c. Unlike the previous subsection, the regulation here increases the radial distributions of the linearized $\delta f_n$ corrections for all radii since the bulk viscous corrections are large throughout the entire hypersurface. These results are not so surprising since we expect the linearized $\delta f_n$ approaches to break down when the viscous pressures are large.

For the PTM and PTB distributions, the transition from a linearized $\delta f_n$ correction to a modified equilibrium distribution is delayed until about $\tau \approx 4.25 - 4.5$\,fm/$c$. As a consequence, their particle production rates are more strongly regulated than in the previous scenario; the radial distributions are also modified but only for $r \sim 7.5 - 9$\,fm. In the PTB case, the regulated space-time distributions turn out to be lower than unregulated distributions.\footnote{%
    Although small, this effect is also observed for protons at $\tau < 4 - 5$\,fm/$c$ and $r \sim 7.5$\,fm for the Chapman-Enskog and PTM distributions.} 
This is because the linearized $\delta f_n$ correction \eqref{eq:jonah_linear} strongly softens the particle distributions at low momentum. Furthermore, the linearized renormalization factor $1 + \delta z_\Pi$ increases the overall magnitude of the distribution function. As a result, the linearized PTB correction is regulated mostly by the upper bound $\delta f_n \leq f_{\eq,n}$ in Eq.~\eqref{eq:df_reg}. We recall that this choice for the upper bound, which was used to compute the maximum hadron number, is somewhat arbitrary.\footnote{%
    It is partially motivated by the assumption that hydrodynamics is valid on the hypersurface, which implies that $\delta f_n$ should be smaller in magnitude than $f_{\eq,n}$.} 
One could either raise this upper bound by hand or replace it with a more explicit calculation of $\delta f_{n,\text{max}}$ \cite{Shen:2014vra}. However, we will not pursue this here as the effects are only modest. As far as the sampled space-time distributions are concerned, they are in excellent agreement with the regulated space-time distributions -- the lower subpanels in Fig.~\ref{F_df_space_155} demonstrate that the particle sampler is sampling the particle yields correctly. The unphysical features in the temporal distributions caused by the sudden transition from modified equilibrium to linearized distributions for the PTM and PTB options wherever the former break down illustrate the conceptual issues arising from particlizing fluids with very large dissipative flows. Whenever such features are observed the user should assess the level of trust to be placed in the results based on physics considerations rather than the sampler algorithm itself.

\begin{figure}[!t]
\makebox[\textwidth][c]{\includegraphics[width=\textwidth]{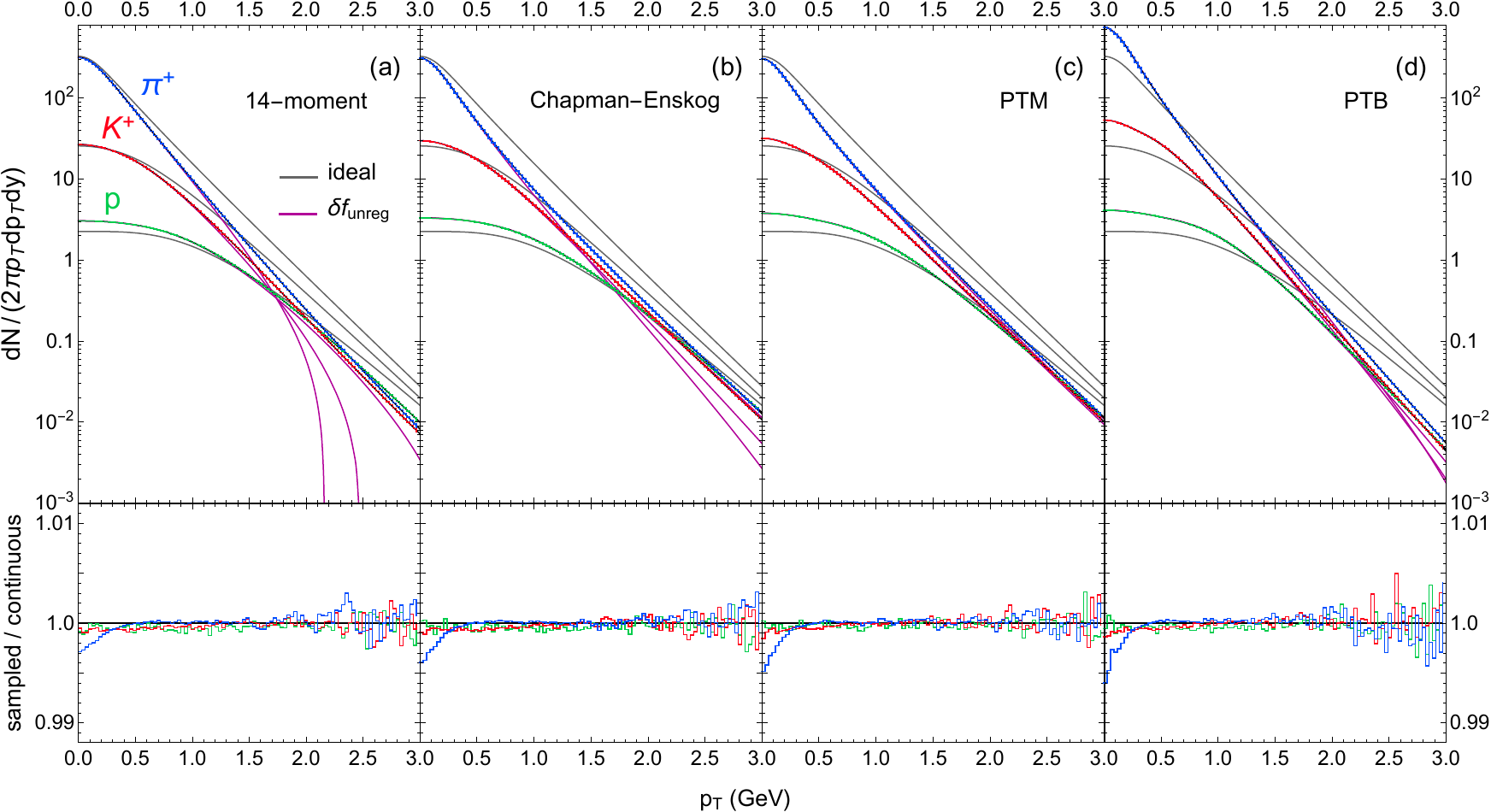}}
\caption{(Color online)
The same comparisons made in Figure~\ref{F_df_space_155} but for the azimuthally-averaged transverse momentum spectra.
\label{F_df_spectra_155}
}
\end{figure}

Figure~\ref{F_df_spectra_155} shows the momentum spectra for this hydrodynamic event. One can see that the spectra of the 14-moment approximation are much more strongly regulated at high $p_T$ than for the event studied in the preceding subsection. The unregulated pion and kaon spectra quickly become negative due to the quadratic momentum dependence of the bulk viscous correction. The Chapman-Enskog spectra are also strongly regulated, although to a somewhat lesser degree since the strength of the bulk viscous correction relative to the thermal distribution grows only linearly with momentum. Out of the four $\delta f_n$ corrections the PTM spectra are the least affected by regulation although there are some modest regulation effects at high $p_T$ resulting from a more frequent breakdown of the modified equilibrium distribution. The regulation has a greater effect on the PTB spectra than the PTM spectra at high $p_T$ since the linearized $\delta f_n$ correction in this case strongly softens the spectra.

Again, we see that the sampled particle spectra agree very well with the regulated continuous spectra. Compared to Fig.~\ref{F_df_spectra_180} in the preceding subsection, the sampling fluctuations are somewhat larger at high $p_T$ since the large viscous corrections lead to fewer particles produced in this region. 

\subsection{Anisotropic flow in non-central Pb-Pb collision}
\label{S6.2}

Turning to non-central collisions, we test in this section the performance of the \I\ sampler for the anisotropic flow coefficients $v_2(p_T)$ and $v_4(p_T)$. Again, we first consider the case of small viscous corrections on the particlization hypersurface (by tuning the peak of the specific bulk viscosity to a safe distance from this surface), followed by a case of large bulk viscous effects on the particlization surface. 

\subsubsection{Small regulated viscous corrections}
\label{S6.2.1}

\begin{figure}[t!]
\makebox[\textwidth][c]{\includegraphics[width=\textwidth]{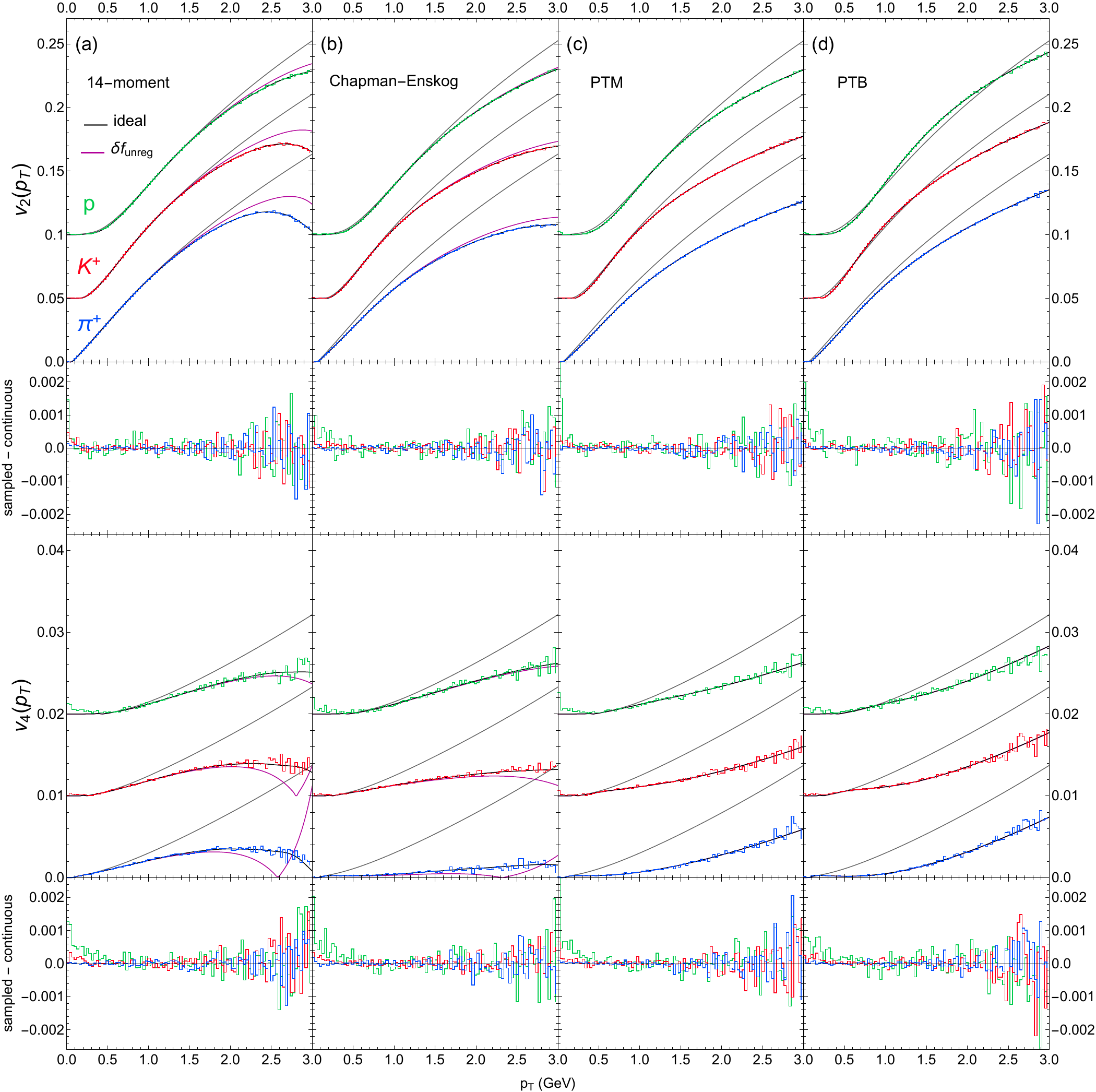}}
\caption{(Color online)
The sampled (solid blue ($\pi^+$), red (K$^+$), and green (p) histograms) and continuous (solid black lines) differential $v_{2,4}(p_T)$ with regulated $\delta f_n$ corrections to the ideal $v_{2,4}(p_T)$ (solid gray lines), for the 2+1d non-central Pb-Pb collision with a peak temperature of $T_p = 180$\,MeV for $\zeta/\mathcal{S}$. For better visibility the results for $v_2$ and $v_4$ for the three particle species are separated vertically by multiples of 0.05 and 0.01, respectively. The $v_{2,4}(p_T)$ with unregulated $\delta f_n$ corrections (solid purple histograms) are also shown. The corresponding lower panels show the difference between the sampled and continuous $v_{2,4}(p_T)$ with regulated $\delta f_n$ corrections.
\label{F_vn_180}
}
\end{figure}

Sampling particles from the non-central Pb-Pb collision with a  peak temperature of $T_p = 180$\,MeV for $\zeta / \mathcal{S}$, we compare in Fig.~\ref{F_vn_180} the sampled $v_2(p_T)$ and $v_4(p_T)$ to those obtained by integrating the Cooper Frye formula with positive-definite integrand, Eq.~(\ref{eq:sampledCFF}). For a given hadron species $j$ the event-averaged discrete anisotropic flow coefficients are computed using the formula
\be
  v_k(p_{T,j}) = \left| \frac{1}{\Delta N_j(p_T)}
  \sum\limits_{n=1}^{\Delta N_j(p_T)} \exp\left(ik\phi_{p,n}\right)\right|
\ee
where $\Delta N_j(p_T)$ is the total number of particles of type $j$ in the transverse momentum bin $[p_{T,j} - \frac{1}{2}\Delta p_T, p_{T,j} + \frac{1}{2}\Delta p_T]$.\footnote{%
    For the purpose of this test we simply sum over all particles sampled from {\em all} events, with $\phi_p$ measured relative to the reaction plane defined by the beam and impact parameter directions, rather than following the experimental procedure of measuring for each event $\phi_p$ relative to the event plane, estimated from the $p_T$-integrated directed or elliptic flow of all charged hadrons.}
The colored histograms in Fig.~\ref{F_vn_180} show the resulting sampled $v_{2}(p_T)$ and $v_{4}(p_T)$ for pions, kaons and protons. The black solid lines show the exact numerical result from the continuous Cooper-Frye formula~\eqref{eq:vn} with both regulated viscous and particle outflow corrections whereas the gray solid lines (labeled \textit{ideal}) use the local equilibrium distribution (without viscous corrections) and outflow correction.

From earlier studies \cite{Schenke:2010rr, Shen:2014lye} it is known that the observed suppression of $v_{2,4}(p_T)$ relative to the \textit{ideal} curves (i.e. $\delta f_n$ = 0) is mainly due to the shear stress correction $\delta f_{\pi,n}$. However, the bulk viscous correction $\delta f_{\Pi,n}$ is also important since it tends to counteract this effect by somewhat increasing $v_{2,4}(p_T)$ at higher $p_T$ \cite{Noronha-Hostler:2013gga, Noronha-Hostler:2015qmd}. In the 14-moment approximation and Chapman-Enskog expansion, the regulation limits the bulk viscous corrections, causing the regulated elliptic flow $v_2(p_T)$ to fall at high $p_T$, but the regulation of the shear stress correction partially cancels this effect. In contrast, the regulation causes the quadrangular flow $v_4(p_T)$ to increase slightly at high $p_T$ since it is more sensitive to the regulation of the shear stress correction than the bulk viscous correction. The PTM and PTB predictions for $v_{2,4}(p_T)$ are not affected at all by the regulation since the transition to a modified equilibrium distribution occurs very early and the few particles produced before this time have not yet developed any anisotropic flow.

Overall, the sampled $v_{2,4}(p_T)$ are in very good agreement with the regulated continuous $v_{2,4}(p_T)$. The larger fluctuations of the sampled $v_{2,4}(p_T)$ in the low and high $p_T$ regions are of statistical nature since there are fewer particles in these $p_T$ bins. In particular, with the numbers of particles sampled for this test, the statistical fluctuations of the sampled quadrangular flow $v_4(p_T)$ are of similar magnitude as its mean value which here is about an order of magnitude smaller than the elliptic flow $v_2(p_T)$, due to the smooth geometric hydrodynamic profile studied in this work. This illustrates the increasing statistical demands associated with measurements of higher-order anisotropic flow coefficients from sampled particle distributions.
%
\begin{figure}[!t]
\makebox[\textwidth][c]{\includegraphics[width=\textwidth]{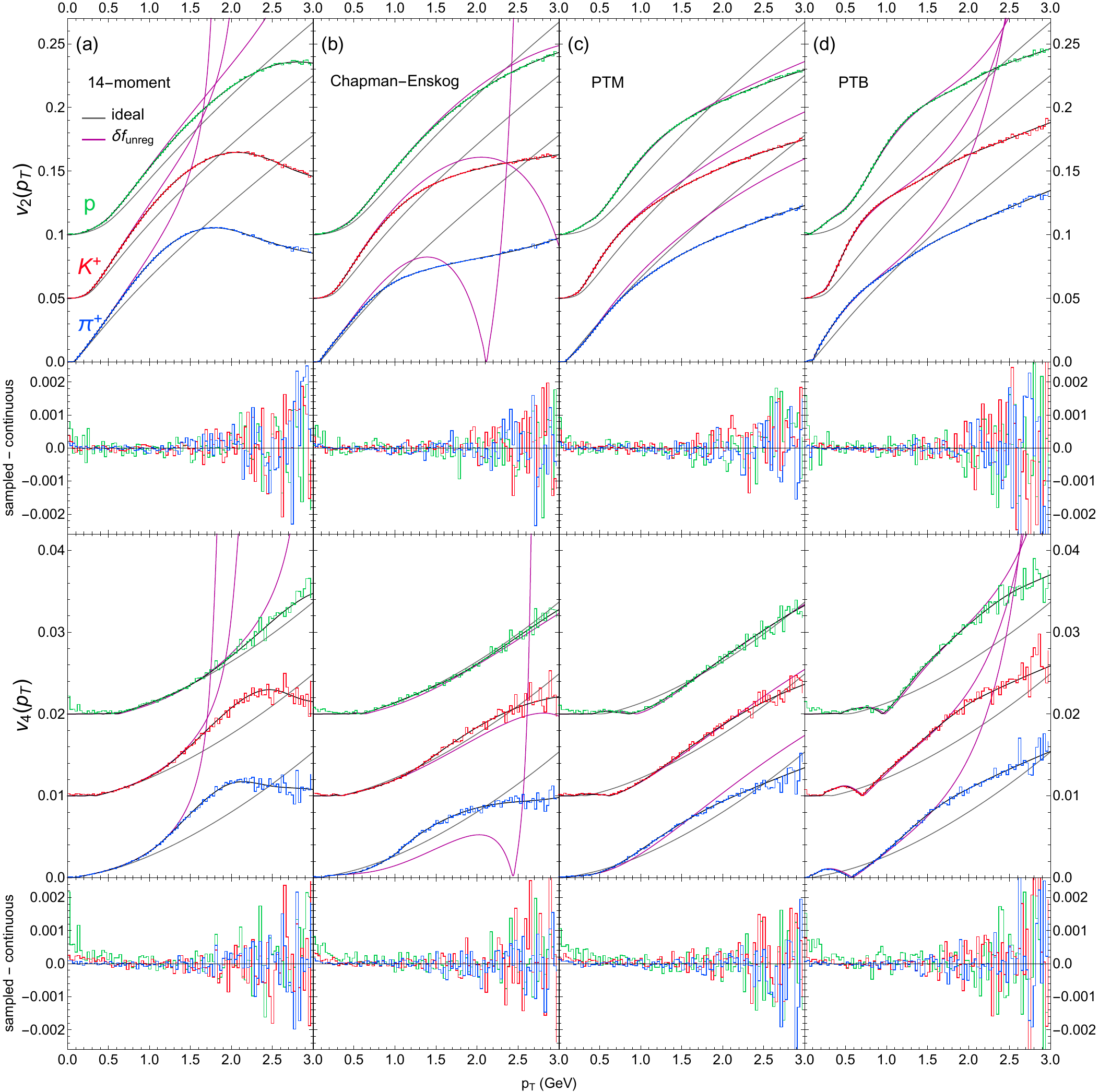}}
\caption{(Color online)
Same as Fig.~\ref{F_vn_180} but for the non-central Pb-Pb collision with a peak temperature of $T_p = 155$\,MeV for $\zeta/\mathcal{S}$.
\label{F_vn_155}
}
\end{figure}
%
\subsubsection{Large regulated viscous corrections}
\label{S6.2.2}
Figure~\ref{F_vn_155} shows what happens when the bulk viscous correction effects on the particlization surface are increased by moving the peak of $\zeta / \mathcal{S}$ to $T_p = 155$\,MeV, i.e. closer to the particlization temperature of 150\,MeV. Clearly the bulk viscous corrections now have a considerably stronger impact on the $v_{2,4}(p_T)$ than in the preceding subsection. Nevertheless, we again find excellent agreement between the sampled and continuous regulated elliptic and quadrangalar flows $v_{2,4}(p_T)$. At high $p_T$ the statistical fluctuations are considerably larger than in the preceding subsection, as a result of the stronger softening of the spectra by the larger bulk viscous pressure effects. Obviously, this could be addressed by sampling more events, at a numerical cost.

\subsection{Benchmarks}
\label{S6.3}

We close this section by presenting benchmarks for the time needed to sample different numbers of particles or events from the boost-invariant (2+1)-dimensional hypersurfaces used for the tests presented in this paper. For these benchmark tests the sampling routine was executed on a single-core Intel Xeon E5-2680 v4 CPU with the G++ compiler and $-03$ optimization.

\begin{figure}[t]
\makebox[\textwidth][c]{\includegraphics[width=\textwidth]{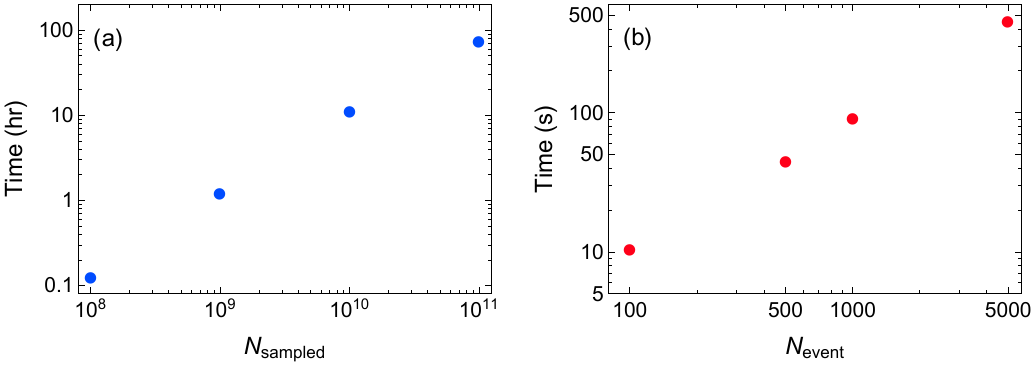}}
\caption{(Color online)
Typical time needed to sample a fixed number of particles (a) or events (b) from a longitudinally boost-invariant (2+1)-d hypersurface. In panel (a), we average the time over both the central and non-central collisions and $\delta f_n$ corrections. In panel (b), we take the average over the central collisions and $\delta f_n$ corrections. 
\label{F_bench}
}
\end{figure}

Figure~\ref{F_bench}a shows the average times needed to sample $N_\text{sampled} =$ 10$^8$, 10$^9$, 10$^{10}$ and 10$^{11}$ particles from a given hydrodynamic hypersurface. The largest of these samples took about 74 hours to generate; this is the sample used for the comparison tests presented earlier in this Section. Such large samples are only needed for precision tests of the sampler; for real applications, it is not practical to sample millions of events from each individual hypersurface. Instead, realistic model-to-data comparisons must generate particles from a sufficiently large number of hydrodynamic events with fluctuating initial conditions, to properly sample the unavoidable initial-state quantum fluctuations. In practice, one typically samples around a thousand particlization events per hydrodynamic simulation. If one is only interested in experimental observables in the mid-rapidity region, one may use a smaller rapidity window as well. For the benchmark test shown in Fig.~\ref{F_bench}b we set $y_{\text{max}} = 3$. The figure shows the average time to sample $N_\mathrm{event}=100,\ 500,\, 1000$\, and 5000 events for central Pb-Pb collision surfaces; the typical time needed to sample 1000 events and write the particle lists to file is about 90$s$.

While sampling the (2+1)-d hypersurfaces is quite fast, it will take somewhat longer to sample (3+1)-d hypersurfaces without boost-invariance because in that case the Poisson distributions for the number of particles emitted from freeze-out cells at different space-time rapidities must now be sampled independently. It may be possible to accelerate the particle sampler by parallelizing the freezeout cells, just like we did for the continuous Cooper-Frye formula. However, this would require restructuring the sampling algorithm, which has been designed to run efficiently on a single-core CPU. We leave this development to future work. 

\section{Summary and Outlook}
\label{S7}

In this work we have documented the performance of a particlization module that is capable of sampling particles from the Cooper-Frye formula with several different forms of the dissipative correction $\delta f_n$ to the distribution function on the particlization hypersurface. This code provides the community with extended abilities for exploring the bulk and shear viscous effects on experimental observables in a realistic setting where finite numbers of particles are Monte-Carlo sampled from fluctuating heavy-ion collision events, instead of computing smooth averages from the Cooper-Frye integral (corresponding to infinite sampling statistics). This provides necessary access to studying the model sensitivity of dissipative effects at particlization on event-by-event fluctuating experimental observables.

The quantitative precision and reliability of our particle sampler has been demonstrated with a number of tests using two typical hydrodynamic event surfaces from (2+1)-dimensional relativistic dissipative fluid dynamical simulations with longitudinal boost-invariance, one for central and one for non-central collisions. The simplifying assumption of boost-invariance has no influence on the precision of the sampler but slightly decreases the time needed to generate the desired event or particle statistics. The \I\ code itself works equally well for particlization surfaces without boost-invariance. It is also designed for hydrodynamic codes that propagate net baryon number and baryon diffusion current. However, these additional features, while implemented, have not yet been tested. Corresponding tests will be performed and reported in future work.

\section{Acknowledgements}
\label{S8}

The authors would like to thank Matthew Golden for assisting in the early stages of this project, in particular for writing the first version of the parallelized GPU code for the continuous Cooper-Frye spectra. This work was supported by the National Science Foundation (NSF) within the framework of the JETSCAPE Collaboration under Award No. \rm{ACI-1550223}. Additional partial support by the U.S. Department of Energy (DOE), Office of Science, Office for Nuclear Physics under Award No. \rm{DE-SC0004286} and within the framework of the BEST and JET Collaborations is also acknowledged. 

\begin{appendices}
\section{Coefficients of the viscous corrections}
\label{App_A}

Here we list the coefficients of the four $\delta f_n$ corrections in this paper. For a more detailed derivation of these coefficients see Ref.~\cite{feqmod}. The \I\ code can generate the $\delta f_n$ coefficients for a given hadron resonance gas. In this paper, we  compute the $\delta f_n$ coefficients with the hadron species included in SMASH \cite{Weil:2016zrk} but this resonance table can be swapped out for the one used in URQMD \cite{Bass:1998ca} or for any other such table. 

\subsection{14-moment approximation}
\label{appa1}

The coefficients of the 14-moment approximation~\eqref{eq:14} are
\bs
\beal
A_T &= \frac{\mathcal{P}}{\A_{21}\mathcal{P} + \N_{31}\mathcal{Q} + \J_{41}\mathcal{R}} \,,\\
A_B &= \frac{\mathcal{Q}}{\A_{21}\mathcal{P} + \N_{31}\mathcal{Q} + \J_{41}\mathcal{R}} \,,\\
A_E &= \frac{\mathcal{R}}{\A_{21}\mathcal{P} + \N_{31}\mathcal{Q} + \J_{41}\mathcal{R}} \,,\\
A_V & = \frac{\J_{41}}{\N^2_{31} - \mathcal{M}_{21} \J_{41}} \,,\\
A_Q & = -\frac{\N_{31}}{\N^2_{31} - \mathcal{M}_{21} \J_{41}} \,,\\
A_\pi &= \frac{1}{2\left(\ene{+}\Peq\right)T^2}\,,
\end{align}
\es
where $\mathcal{E}$ is the energy density, $\mathcal{P}_\eq$ is the equilibrium pressure and
\bs
\beal
\mathcal{P} = &\, \N_{30}^2 - \J_{40}\mathcal{M}_{20} \,,\\
\mathcal{Q} = &\, \B_{10}\J_{40} {-} \A_{20}\N_{30} \,,\\
\mathcal{R} = &\, \A_{20}\mathcal{M}_{20} - \B_{10}\N_{30} \,.
\end{align}
\es
The thermodynamic integrals $\mathcal{J}_{rq}$, $\mathcal{N}_{rq}$, $\mathcal{M}_{rq}$, $\mathcal{A}_{rq}$ and $\mathcal{B}_{rq}$ are defined as
\bs
\beal
\mathcal{J}_{rq} &= \frac{1}{(2q{+}1)!!}\sum_n \int_p (u \cdot p)^{r-2q} (- p \cdot \Delta \cdot p)^{q} \,f_{\eq,n} \bar{f}_{\eq,n} \,,\\
\mathcal{N}_{rq} &= \frac{1}{(2q{+}1)!!}\sum_n b_n \int_p (u \cdot p)^{r-2q} (- p \cdot \Delta \cdot p)^{q} \,f_{\eq,n} \bar{f}_{\eq,n} \,,\\
\mathcal{M}_{rq} &= \frac{1}{(2q{+}1)!!}\sum_n b_n^2 \int_p (u \cdot p)^{r-2q} (- p \cdot \Delta \cdot p)^{q} \,f_{\eq,n}\bar{f}_{\eq,n} \,,\\
\mathcal{A}_{rq} &= \frac{1}{(2q{+}1)!!}\sum_n m_n^2 \int_p (u \cdot p)^{r-2q} (- p \cdot \Delta \cdot p)^{q} \,f_{\eq,n}\bar{f}_{\eq,n} \,,\\
\mathcal{B}_{rq} &= \frac{1}{(2q{+}1)!!}\sum_n b_n m_n^2 \int_p (u \cdot p)^{r-2q} (- p \cdot \Delta \cdot p)^{q} \,f_{\eq,n}\bar{f}_{\eq,n} \,.
\end{align}
\es

\subsection{Chapman-Enskog expansion}
\label{appa2}

The coefficients of the Chapman Enskog expansion~\eqref{eq:CE} are
\bs
\beal
\mathcal{G} &= T\left(\frac{(\ene{+}\Peq)\N_{20} - n_B \J_{30}}{\J_{30} \mathcal{M}_{10} - \N_{20}^2}\right) \,,\\
\mathcal{F} &= T^2\left(\frac{n_B\N_{20} - (\ene{+}\Peq)\mathcal{M}_{10}}{\J_{30} \mathcal{M}_{10} - \N_{20}^2}\right)  \,,\\ 
\beta_\Pi &= \mathcal{G} n_B T + \frac{\mathcal{F}\left(\ene{+}\Peq \right)}{T} + \frac{5\mathcal{J}_{32}}{3T}\,,\\
\beta_V &= \mathcal{M}_{11} - \frac{n_B^2 T}{\ene{+}\Peq} \,,\\
\beta_\pi &= \frac{\mathcal{J}_{32}}{T}\,,
\end{align}
\es
where $n_B$ is the net baryon density. One notes that the PTM distribution uses the same coefficients to modify the local-equilibrium distribution. 

\subsection{Pratt-Torrieri-Bernhard distribution}
\label{appa3}

The parameters $z_\Pi$ and $\lambda_\Pi$ are computed from the formula
\bs
\beal
    \mathcal{Z}_\Pi &= \frac{\mathcal{E}}{\mathcal{L}_{20}} \,,
\\
    \Pi &= \mathcal{Z}_\Pi \mathcal{L}_{21} - \mathcal{P}_\eq \,,
\end{align}
\es
where $\mathcal{L}_{rq}$ is defined as
\be
    \mathcal{L}_{rq} = \frac{1}{(2q+1)!!}\sum_n \int_p 
    (u \cdot p)^{r-2q} (- p \cdot \Delta \cdot p)^{q} \,f_{\lambda,n}\,,
\ee
with
\be
    f_{\lambda,n} = \frac{g_n}{(1+\lambda_\Pi)^3} \left[\exp\left( 
      \frac{1}{T}\sqrt{m_n^2 
    - \frac{p\cdot \Delta\cdot p}
           {(1{+}\lambda_\Pi)^2}} \right) + \Theta_n\right]^{-1}\,.
\ee
For a given value of $\lambda_\Pi$, one can compute the corresponding outputs for $\mathcal{Z}_\Pi$ and $\Pi$ to tabulate the parameters as a function of $\Pi$. In the limit of small bulk pressure the parameters linearize to
\bs
\beal
    1 + \delta\mathcal{Z}_\Pi 
    &= 1 - \frac{3\Pi \Peq}{5 \beta_\pi \ene - 3\Peq(\ene{+}\Peq)}, 
\\
    \delta\lambda_\Pi 
    &= \frac{\Pi \ene}{5 \beta_\pi \ene - 3\Peq(\ene{+}\Peq)}.
\end{align}
\es

\end{appendices}

\bibliographystyle{elsarticle-num}
\bibliography{iS3D}

\begin{thebibliography}{10}
\expandafter\ifx\csname url\endcsname\relax
  \def\url#1{\texttt{#1}}\fi
\expandafter\ifx\csname urlprefix\endcsname\relax\def\urlprefix{URL }\fi
\expandafter\ifx\csname href\endcsname\relax
  \def\href#1#2{#2} \def\path#1{#1}\fi

\bibitem{PhysRevD.10.186}
F.~Cooper, G.~Frye, Single-particle distribution in the hydrodynamic and
  statistical thermodynamic models of multiparticle production, Phys. Rev. D 10
  (1974) 186--189.
\newblock \href {http://dx.doi.org/10.1103/PhysRevD.10.186}
  {\path{doi:10.1103/PhysRevD.10.186}}.

\bibitem{Akamatsu:2018olk}
Y.~Akamatsu, et~al., {Dynamically integrated transport approach for heavy-ion
  collisions at high baryon density}, Phys. Rev. C98 (2018) 024909.
\newblock \href {http://arxiv.org/abs/1805.09024} {\path{arXiv:1805.09024}},
  \href {http://dx.doi.org/10.1103/PhysRevC.98.024909}
  {\path{doi:10.1103/PhysRevC.98.024909}}.

\bibitem{Pierog:2013ria}
T.~Pierog, I.~Karpenko, J.~M. Katzy, E.~Yatsenko, K.~Werner, {EPOS LHC: Test of
  collective hadronization with data measured at the CERN Large Hadron
  Collider}, Phys. Rev. C92 (2015) 034906.
\newblock \href {http://arxiv.org/abs/1306.0121} {\path{arXiv:1306.0121}},
  \href {http://dx.doi.org/10.1103/PhysRevC.92.034906}
  {\path{doi:10.1103/PhysRevC.92.034906}}.

\bibitem{Molnar:2014fva}
D.~Molnar, Z.~Wolff, Self-consistent conversion of a viscous fluid to
  particles, Phys. Rev. C95 (2017) 024903.
\newblock \href {http://arxiv.org/abs/1404.7850} {\path{arXiv:1404.7850}},
  \href {http://dx.doi.org/10.1103/PhysRevC.95.024903}
  {\path{doi:10.1103/PhysRevC.95.024903}}.

\bibitem{Wolff:2016vcm}
Z.~Wolff, D.~Molnar, {Flow harmonics from self-consistent particlization of a
  viscous fluid}, Phys. Rev. C96~(4) (2017) 044909.
\newblock \href {http://arxiv.org/abs/1611.09185} {\path{arXiv:1611.09185}},
  \href {http://dx.doi.org/10.1103/PhysRevC.96.044909}
  {\path{doi:10.1103/PhysRevC.96.044909}}.

\bibitem{Damodaran:2017ior}
M.~Damodaran, D.~Molnar, G.~G. Barnaföldi, D.~Berényi, M.~Ferenc Nagy-Egri,
  {Testing and improving shear viscous phase space correction models}\href
  {http://arxiv.org/abs/1707.00793} {\path{arXiv:1707.00793}}.

\bibitem{CPA:CPA3160020403}
H.~Grad, On the kinetic theory of rarefied gases, Communications on Pure and
  Applied Mathematics 2~(4) (1949) 331--407.
\newblock \href {http://dx.doi.org/10.1002/cpa.3160020403}
  {\path{doi:10.1002/cpa.3160020403}}.

\bibitem{chapman1990mathematical}
S.~Chapman, T.~Cowling, D.~Burnett, C.~Cercignani, The Mathematical Theory of
  Non-uniform Gases: An Account of the Kinetic Theory of Viscosity, Thermal
  Conduction and Diffusion in Gases, Cambridge Mathematical Library, Cambridge
  University Press, 1990.

\bibitem{Pratt:2010jt}
S.~Pratt, G.~Torrieri, {Coupling Relativistic Viscous Hydrodynamics to
  Boltzmann Descriptions}, Phys. Rev. C82 (2010) 044901.
\newblock \href {http://arxiv.org/abs/1003.0413} {\path{arXiv:1003.0413}},
  \href {http://dx.doi.org/10.1103/PhysRevC.82.044901}
  {\path{doi:10.1103/PhysRevC.82.044901}}.

\bibitem{Bernhard:2018hnz}
J.~E. Bernhard, {Bayesian parameter estimation for relativistic heavy-ion
  collisions}, Ph.D. thesis, Duke University (2018-04-19).
\newblock \href {http://arxiv.org/abs/1804.06469} {\path{arXiv:1804.06469}}.

\bibitem{feqmod}
M.~McNelis, D.~Everett, U.~Heinz, {Modified equilibrium distributions for
  Cooper-Frye particlization}, {} {} (2020) {}, in preparation.
\newblock \href {http://dx.doi.org/{}} {\path{doi:{}}}.

\bibitem{Bernhard:2016tnd}
J.~E. Bernhard, J.~S. Moreland, S.~A. Bass, J.~Liu, U.~Heinz, {Applying
  Bayesian parameter estimation to relativistic heavy-ion collisions:
  simultaneous characterization of the initial state and quark-gluon plasma
  medium}, Phys. Rev. C94 (2016) 024907.
\newblock \href {http://arxiv.org/abs/1605.03954} {\path{arXiv:1605.03954}},
  \href {http://dx.doi.org/10.1103/PhysRevC.94.024907}
  {\path{doi:10.1103/PhysRevC.94.024907}}.

\bibitem{Putschke:2019yrg}
J.~H. Putschke, et~al., {The JETSCAPE framework. }\href
  {http://arxiv.org/abs/1903.07706} {\path{arXiv:1903.07706}}.

\bibitem{Schenke:2010nt}
B.~Schenke, S.~Jeon, C.~Gale, {(3+1)D hydrodynamic simulation of relativistic
  heavy-ion collisions}, Phys. Rev. C82 (2010) 014903.
\newblock \href {http://arxiv.org/abs/1004.1408} {\path{arXiv:1004.1408}},
  \href {http://dx.doi.org/10.1103/PhysRevC.82.014903}
  {\path{doi:10.1103/PhysRevC.82.014903}}.

\bibitem{Denicol:2012cn}
G.~S. Denicol, H.~Niemi, E.~Molnar, D.~H. Rischke, {Derivation of transient
  relativistic fluid dynamics from the Boltzmann equation}, Phys. Rev. D85
  (2012) 114047, [Erratum: Phys. Rev.D 91 (2015) 039902].
\newblock \href {http://arxiv.org/abs/1202.4551} {\path{arXiv:1202.4551}},
  \href {http://dx.doi.org/10.1103/PhysRevD.85.114047}
  {\path{doi:10.1103/PhysRevD.85.114047}}.

\bibitem{Shen:2014vra}
C.~Shen, Z.~Qiu, H.~Song, J.~Bernhard, S.~Bass, U.~Heinz, {The iEBE-VISHNU code
  package for relativistic heavy-ion collisions}, Comput. Phys. Commun. 199
  (2016) 61--85.
\newblock \href {http://arxiv.org/abs/1409.8164} {\path{arXiv:1409.8164}},
  \href {http://dx.doi.org/10.1016/j.cpc.2015.08.039}
  {\path{doi:10.1016/j.cpc.2015.08.039}}.

\bibitem{Pang:2018zzo}
L.-G. Pang, H.~Petersen, X.-N. Wang, {Pseudorapidity distribution and
  decorrelation of anisotropic flow within the open-computing-language
  implementation CLVisc hydrodynamics}, Phys. Rev. C97 (2018) 064918.
\newblock \href {http://arxiv.org/abs/1802.04449} {\path{arXiv:1802.04449}},
  \href {http://dx.doi.org/10.1103/PhysRevC.97.064918}
  {\path{doi:10.1103/PhysRevC.97.064918}}.

\bibitem{Bass:1998ca}
S.~A. Bass, et~al., {Microscopic models for ultrarelativistic heavy ion
  collisions}, Prog. Part. Nucl. Phys. 41 (1998) 255--369, [Prog. Part. Nucl.
  Phys.41,225(1998)].
\newblock \href {http://arxiv.org/abs/nucl-th/9803035}
  {\path{arXiv:nucl-th/9803035}}, \href
  {http://dx.doi.org/10.1016/S0146-6410(98)00058-1}
  {\path{doi:10.1016/S0146-6410(98)00058-1}}.

\bibitem{Weil:2016zrk}
J.~Weil, et~al., {Particle production and equilibrium properties within a new
  hadron transport approach for heavy-ion collisions}, Phys. Rev. C94 (2016)
  054905.
\newblock \href {http://arxiv.org/abs/1606.06642} {\path{arXiv:1606.06642}},
  \href {http://dx.doi.org/10.1103/PhysRevC.94.054905}
  {\path{doi:10.1103/PhysRevC.94.054905}}.

\bibitem{Romatschke:2017ejr}
P.~Romatschke, U.~Romatschke, {Relativistic Fluid Dynamics In and Out of
  Equilibrium}, Cambridge Monographs on Mathematical Physics, Cambridge
  University Press, 2019.
\newblock \href {http://arxiv.org/abs/1712.05815} {\path{arXiv:1712.05815}},
  \href {http://dx.doi.org/10.1017/9781108651998}
  {\path{doi:10.1017/9781108651998}}.

\bibitem{Heinz:2019dbd}
U.~Heinz, J.~S. Moreland, {Hydrodynamic flow in small systems or: “How the
  heck is it possible that a system emitting only a dozen particles can be
  described by fluid dynamics?”}, J. Phys. Conf. Ser. 1271 (2019) 012018.
\newblock \href {http://arxiv.org/abs/1904.06592} {\path{arXiv:1904.06592}},
  \href {http://dx.doi.org/10.1088/1742-6596/1271/1/012018}
  {\path{doi:10.1088/1742-6596/1271/1/012018}}.

\bibitem{Jaiswal:2019cju}
S.~Jaiswal, C.~Chattopadhyay, A.~Jaiswal, S.~Pal, U.~Heinz, {Exact solutions
  and attractors of higher-order viscous fluid dynamics for Bjorken flow},
  Phys. Rev. C100 (2019) 034901.
\newblock \href {http://arxiv.org/abs/1907.07965} {\path{arXiv:1907.07965}},
  \href {http://dx.doi.org/10.1103/PhysRevC.100.034901}
  {\path{doi:10.1103/PhysRevC.100.034901}}.

\bibitem{Israel:1976tn}
W.~Israel, {Nonstationary irreversible thermodynamics: A Causal relativistic
  theory}, Annals Phys. 100 (1976) 310--331.
\newblock \href {http://dx.doi.org/10.1016/0003-4916(76)90064-6}
  {\path{doi:10.1016/0003-4916(76)90064-6}}.

\bibitem{Israel:1979wp}
W.~Israel, J.~M. Stewart, {Transient relativistic thermodynamics and kinetic
  theory}, Annals Phys. 118 (1979) 341--372.
\newblock \href {http://dx.doi.org/10.1016/0003-4916(79)90130-1}
  {\path{doi:10.1016/0003-4916(79)90130-1}}.

\bibitem{Monnai:2009ad}
A.~Monnai, T.~Hirano, {Effects of bulk viscosity at freezeout}, Phys. Rev. C80
  (2009) 054906.
\newblock \href {http://arxiv.org/abs/0903.4436} {\path{arXiv:0903.4436}},
  \href {http://dx.doi.org/10.1103/PhysRevC.80.054906}
  {\path{doi:10.1103/PhysRevC.80.054906}}.

\bibitem{Anderson_Witting_1974}
J.~Anderson, H.~Witting, {A relativistic relaxation-time model for the
  Boltzmann equation}, Physica 74 (1974) 466--488.

\bibitem{Jaiswal:2014isa}
A.~Jaiswal, R.~Ryblewski, M.~Strickland, {Transport coefficients for bulk
  viscous evolution in the relaxation time approximation}, Phys. Rev. C90
  (2014) 044908.
\newblock \href {http://arxiv.org/abs/1407.7231} {\path{arXiv:1407.7231}},
  \href {http://dx.doi.org/10.1103/PhysRevC.90.044908}
  {\path{doi:10.1103/PhysRevC.90.044908}}.

\bibitem{Bazow:2016yra}
D.~Bazow, U.~Heinz, M.~Strickland, {Massively parallel simulations of
  relativistic fluid dynamics on graphics processing units with CUDA}, Comput.
  Phys. Commun. 225 (2018) 92--113.
\newblock \href {http://arxiv.org/abs/1608.06577} {\path{arXiv:1608.06577}},
  \href {http://dx.doi.org/10.1016/j.cpc.2017.01.015}
  {\path{doi:10.1016/j.cpc.2017.01.015}}.

\bibitem{Miller:2007ri}
M.~L. Miller, K.~Reygers, S.~J. Sanders, P.~Steinberg, {Glauber modeling in
  high energy nuclear collisions}, Ann. Rev. Nucl. Part. Sci. 57 (2007)
  205--243.
\newblock \href {http://arxiv.org/abs/nucl-ex/0701025}
  {\path{arXiv:nucl-ex/0701025}}, \href
  {http://dx.doi.org/10.1146/annurev.nucl.57.090506.123020}
  {\path{doi:10.1146/annurev.nucl.57.090506.123020}}.

\bibitem{Denicol:2009am}
G.~S. Denicol, T.~Kodama, T.~Koide, P.~Mota, {Effect of bulk viscosity on
  Elliptic Flow near QCD phase transition}, Phys. Rev. C80 (2009) 064901.
\newblock \href {http://arxiv.org/abs/0903.3595} {\path{arXiv:0903.3595}},
  \href {http://dx.doi.org/10.1103/PhysRevC.80.064901}
  {\path{doi:10.1103/PhysRevC.80.064901}}.

\bibitem{Du:2019obx}
L.~Du, U.~Heinz, {(3+1)-dimensional dissipative relativistic fluid dynamics at
  non-zero net baryon density, }\href {http://arxiv.org/abs/1906.11181}
  {\path{arXiv:1906.11181}}, \href
  {http://dx.doi.org/10.1016/j.cpc.2019.107090}
  {\path{doi:10.1016/j.cpc.2019.107090}}.

\bibitem{Huovinen:2012is}
P.~Huovinen, H.~Petersen, {Particlization in hybrid models}, Eur. Phys. J. A48
  (2012) 171.
\newblock \href {http://arxiv.org/abs/1206.3371} {\path{arXiv:1206.3371}},
  \href {http://dx.doi.org/10.1140/epja/i2012-12171-9}
  {\path{doi:10.1140/epja/i2012-12171-9}}.

\bibitem{Borsanyi:2010cj}
S.~Borsanyi, G.~Endrodi, Z.~Fodor, A.~Jakovac, S.~D. Katz, S.~Krieg, C.~Ratti,
  K.~K. Szabo, {The QCD equation of state with dynamical quarks}, JHEP 11
  (2010) 077.
\newblock \href {http://arxiv.org/abs/1007.2580} {\path{arXiv:1007.2580}},
  \href {http://dx.doi.org/10.1007/JHEP11(2010)077}
  {\path{doi:10.1007/JHEP11(2010)077}}.

\bibitem{Borsanyi:2013bia}
S.~Borsanyi, Z.~Fodor, C.~Hoelbling, S.~D. Katz, S.~Krieg, K.~K. Szabo, {Full
  result for the QCD equation of state with 2+1 flavors}, Phys. Lett. B730
  (2014) 99--104.
\newblock \href {http://arxiv.org/abs/1309.5258} {\path{arXiv:1309.5258}},
  \href {http://dx.doi.org/10.1016/j.physletb.2014.01.007}
  {\path{doi:10.1016/j.physletb.2014.01.007}}.

\bibitem{Bazavov:2014pvz}
A.~Bazavov, et~al., {Equation of state in ( 2+1 )-flavor QCD}, Phys. Rev. D90
  (2014) 094503.
\newblock \href {http://arxiv.org/abs/1407.6387} {\path{arXiv:1407.6387}},
  \href {http://dx.doi.org/10.1103/PhysRevD.90.094503}
  {\path{doi:10.1103/PhysRevD.90.094503}}.

\bibitem{Shen:2014lye}
C.~Shen, \href{http://rave.ohiolink.edu/etdc/view?acc_num=osu1405931790}{{The
  standard model for relativistic heavy-ion collisions and electromagnetic
  tomography}}, Ph.D. thesis, Ohio State U. (2014-07-25).
\newline\urlprefix\url{http://rave.ohiolink.edu/etdc/view?acc_num=osu1405931790}

\bibitem{Pratt:2014vja}
S.~Pratt, {Accounting for backflow in hydrodynamic-Boltzmann interfaces}, Phys.
  Rev. C89 (2014) 024910.
\newblock \href {http://arxiv.org/abs/1401.0316} {\path{arXiv:1401.0316}},
  \href {http://dx.doi.org/10.1103/PhysRevC.89.024910}
  {\path{doi:10.1103/PhysRevC.89.024910}}.

\bibitem{OSCAR}
\href{https://karman.physics.purdue.edu/OSCAR-old/docs/file/cascade_output_format/osc99a.html}{{OSCAR}
  {F}ile {F}ormat} (1999).
\newline\urlprefix\url{https://karman.physics.purdue.edu/OSCAR-old/docs/file/cascade_output_format/osc99a.html}

\bibitem{Schenke:2010rr}
B.~Schenke, S.~Jeon, C.~Gale, {Elliptic and triangular flow in event-by-event
  (3+1)D viscous hydrodynamics}, Phys. Rev. Lett. 106 (2011) 042301.
\newblock \href {http://arxiv.org/abs/1009.3244} {\path{arXiv:1009.3244}},
  \href {http://dx.doi.org/10.1103/PhysRevLett.106.042301}
  {\path{doi:10.1103/PhysRevLett.106.042301}}.

\bibitem{Noronha-Hostler:2013gga}
J.~Noronha-Hostler, G.~S. Denicol, J.~Noronha, R.~P.~G. Andrade, F.~Grassi,
  {Bulk viscosity effects in event-by-event relativistic hydrodynamics}, Phys.
  Rev. C88~(4) (2013) 044916.
\newblock \href {http://arxiv.org/abs/1305.1981} {\path{arXiv:1305.1981}},
  \href {http://dx.doi.org/10.1103/PhysRevC.88.044916}
  {\path{doi:10.1103/PhysRevC.88.044916}}.

\bibitem{Noronha-Hostler:2015qmd}
J.~Noronha-Hostler, {Extracting shear viscosity of the Quark Gluon Plasma in
  the presence of bulk viscosity}, in: {Proceedings, 12th Conference on the
  Intersections of Particle and Nuclear Physics (CIPANP 2015): Vail, Colorado,
  USA, May 19-24, 2015}, 2015.
\newblock \href {http://arxiv.org/abs/1512.06315} {\path{arXiv:1512.06315}}.

\end{thebibliography}
\end{document}